\documentclass[12pt]{article}
\usepackage{geometry}
\usepackage{authblk}
\usepackage{graphicx}
\usepackage{url}
\usepackage{booktabs}
\usepackage{array}
\usepackage{multirow}
\usepackage{amsmath}
\usepackage{caption}
\usepackage{color}
\usepackage{float}
\usepackage{adjustbox}
\usepackage{epstopdf}
\providecommand{\keywords}[1]{\vspace{1em}\noindent\textbf{Keywords: }#1}
\usepackage[square,numbers]{natbib}
\newcolumntype{L}[1]{>{\raggedright\arraybackslash}p{#1}}
\newcolumntype{C}[1]{>{\centering\arraybackslash}p{#1}}
\newcolumntype{R}[1]{>{\raggedleft\arraybackslash}p{#1}}

\title{Energy Rates Due to Weak Decay Rates of Vanadium Isotopes in Stellar Environment}
\author[1]{Ramoona Shehzadi}
\author[2]{Jameel-Un Nabi}
\author[2]{Huma Ali}
\affil[1]{Department of Physics, University of the Punjab, Lahore, Pakistan\\\texttt{ramoona.physics@pu.edu.pk}}
\affil[2]{Faculty of Engineering Sciences (FES), GIK Institute of Engineering Sciences and Technology, Topi 23640, Khyber Pakhtunkhwa, Pakistan}
\date{}  

\begin{document}
	
	\maketitle
	
	\begin{abstract}
		The neutrino cooling and gamma heating rates are considered as an
		important input needed to study the final phases of the evolution of
		high-mass stars. The weak-interaction mediated processes, namely the
		$\beta$-decay and electron capture, significantly change the lepton
		to baryon ratio and accelerate the contraction of the core. The
		emission of resulting neutrinos/antineutrinos tend to cool the
		stellar core. On the other hand gamma rays are produced because of electron capture and $\beta$-decay to excited
		states in daughter nuclei. These gamma rays heat the core and contribute to increase of
		entropy which may cause convection to occur. In the present work,
		the weak-interaction heating and cooling rates on a chain of twenty
		two isotopes of vanadium having mass in the range $43-64$ have been
		estimated using the proton-neutron quasiparticle random phase
		approximation theory. The rates have been computed for the
		temperature ranging from ($10^{7} - 3 \times 10^{10}$)\;K and for the
		density range ($10-10^{11}$)\;g/cm$^{3}$. Our calculated neutrino energy loss
		rates have also been compared with the previously reported rates
		calculated using other theoretical models. At high stellar
		temperatures, our rates are larger by 1-2 orders of magnitude as
		compared to previous results.
	\end{abstract}


\section{Introduction}
\label{sec:intro}

The weak-interaction mediated reactions play an important character
in the star's life cycle, especially during the hydrostatic burning
and late stages of stellar evolution~\citep{Bethe90,Lang03}. The
weak-interaction transformation of protons into neutrons initiates
hydrogen burning in the stellar core. The hydrostatic nuclear
burning process advances progressively through H, He, C, Ne, O and
silicon. After the Si-combustion, the core has developed into
Fe-peak nuclei. When the core's mass goes beyond the Chandersakher
limit of $\sim 1.5$ M$_\odot$, the electron degeneracy pressure is
not enough to combat the force of gravity. It sets up the collapse
of the core with about one-fourth of the speed of
light~\citep{Jose11}. Energetic gamma rays generated by the hot
stellar core initiate disintegration of iron-peak nuclei to free
nucleons and $\alpha$-particles. The onfall of electron capture by
free and bound protons reduces the density of electrons and thus the
destruction of the stellar core expedites. This may start a process
termed as core-collapse supernova explosion.

The collapse dynamics is very sensitive to electron to baryon
fraction (Y$_{e}$) and to the core entropy~\citep{Bethe79}. Weak-
interaction processes including lepton (e$^{-}$ or e$^{+}$) capture
and nuclear $\beta^{\pm}$-decay (lepton emission) are essential in
determining these quantities. Electron capture processes lower
Y$_{e}$ whereas beta decay acts conversely. Initially, when value
of Y$_{e}$ is $\sim$ 0.5, beta decay processes are suppressed by
electron degeneracy, but when nuclei tend to be more neutron rich,
beta decays compete electron captures. These processes also result
in the production of neutrinos and anti-neutrinos. During
precollapse stage, for matter densities up to $\sim
10^{11}$\;g/cm$^{3}$, neutrinos created by e$^{-}$-capture on nuclei
release unimpeded from the star, thereby reducing the temperature
and entropy of the core~\citep{Heger01,Janka07}.
Ref.~\citep{Gupta07} has argued that at these matter densities, the
electron captures to excited states and the ensuing gamma emission,
can discharge up to a factor of ten times extra heat subject to the
core's initial mass and composition. The fraction of energy loss in
a reaction due to emitted neutrinos is substantially decreased. The
electron capture on different nuclei and the associated gamma
heating is also very sensitive to the
temperature~\citep{Giannaka15}. Thus, in addition to neutrino
cooling rates, it is valuable to also compute the competing
gamma-ray heating rates for the nuclei in the iron regime.

In the course of late evolutionary stages of high-mass stars, the
nuclear weak-interaction rates and the related neutrino and
anti-neutrino energy-loss/cooling rates, as well as gamma heating
rates are known to be controlled by charge-exchange reactions, also
named as Gamow-Teller (GT) transitions~\citep{Bethe79}. For stellar
temperatures around 300-800 keV, and densities smaller than $\sim
10^{10}$\;g/cm$^{3}$, electron capture takes place on nuclei having
$A \sim 60$. In this scenario, the electron chemical potential,
$\mu_{e}$, is roughly similar to the nuclear Q-values and thus the
weak-interaction rates are susceptible to the complete structure of
GT transition strength. For still larger temperature and densities,
nuclei having A $>$ 65 become plentiful and $\mu_{e}$ is
considerably greater than Q values. At such high densities, the
capture rates primarily rely on the centroid energy and total GT
strength. The GT properties of fp-shell nuclei are of particular
significance during pre-supernova stages of high-mass stars and also
during the collapse of the core~\citep{Fuller}. Due to intense
conditions of temperatures and densities in the astrophysical
environment, GT transitions from thermally excited levels can occur
in addition to GT transitions from the parent ground level. This
characteristic makes the calculation and measurement of GT strength
a difficult job.

The first detailed computations of astrophysical weak-rates for a
broad dimensions of temperature; $0.01 \le T_{9}\;(\text{K}) \le 30
$ and density $1 \le \log \rho$Y$_{e}$\;(g/cm$^{3}) \le 11$ were
done by Fuller et al~\citep{Fuller}. Their calculations comprised
electron and positron capture rates, $\beta^{\pm}$-decay rates and
the related neutrino/anti-neutrino cooling rates. The rates were
determined for 226 nuclei having mass in the range 21 to 60.
The authors assumed a single GT resonance whose properties were
determined using independent particle model (IPM), augmented by
measured data accessible by that time. Brink hypothesis~\citep{BAH}
was employed to consider GT transitions from parent's excited
levels. The data from later experiments~\citep{Vet89, Alford90,
William90, Alford93} revealed that
compared to the values calculated by IPM, the measured total GT
strength in nuclei were quenched. The data also showed that the GT
strengths are heavily segmented over several states in residual
nuclei. The work of Fuller et. al., was later expanded by
Aufderheide and collaborators~\citep{Aufder94} for further heavy
nuclei having $A$ greater than 60. They also considered the quenching
of the GT strength distributions. Results of (p,n) and (n,p)
experiments~\citep{Kateb94} reported the wrong placement of GT
centroid in the calculations of~\citep{Fuller} and~\citep{Aufder94}.
These findings compelled efforts to the development of microscopic
ways for the estimation of weak rates in the astrophysical domain.

The two most effective microscopic theories which are broadly used
for the reliable determination of stellar weak-rates are the
so-called large-scale shell model (LSSM)~\citep{Lang00} and the
proton-neutron quasi-particle random phase approximation (pn-QRPA)
model~\citep{Nabi99}. Both of these models have their own advantages
and disadvantages. There are two primary advantages of the pn-QRPA
model: (1) a large model space extending to 7 major shells maybe
used for the calculations, which allows handling of any arbitrarily
heavy nuclei. (2) instead of using Brink hypothesis, this model
allows to execute a microscopic, state-by-state computation of
weak-interaction rates. Due to this aspect of the pn-QRPA theory,
the reliability of the weak-rates computed at stellar conditions
enhances. Later studies~\citep{Nabi08}-\citep{Nabi10c} have also
shown that Brink hypothesis is an unsafe approximation to be applied
in stellar calculations.

Initially, Nabi and Klapdor-Kleingrothaus employed the pn-QRPA
theory for an extensive calculation of weak-interaction mediated
rates for a broad domain of stellar densities and
temperatures~\citep{Nabi99, Nabi04}. The calculations were carried
out for 709 nuclei having mass number in the range of $18-100$.
Their calculation included lepton emission and capture rates, the
related neutrino cooling rates, probabilities of emission of
beta-delayed nucleon, the energy rates of these emitted particles
and gamma heating rates. In subsequent studies, for
example~\citep{Nabi07, Nabi08}, these rates were further improved
with the inclusion of latest data from experiments, using refined
algorithms and fine adjustments of the parameters used in the model.
A detailed evaluation of the validity and accuracy of the pn-QRPA
estimated weak-rates could be found in Ref.~\citep{Nabi04}.

In this paper, a thorough analysis of the pn-QRPA estimated neutrino
cooling rates and competing gamma heating rates as a result of
weak-interaction processes on vanadium isotopes in the stellar
environment is presented. As per studies of Aufderheide and his
collaborators~\citep{Aufder94}, eight isotopes of vanadium
($^{50-56,58}$V) were enlisted as crucial beta decay and electron
capture nuclei which significantly affect Y$_{e}$ in the final
phases of progression of high-mass stars. The results of GT strength
distributions and corresponding electron capture rates on $^{50}$V
and $^{51}$V nuclide in astrophysical atmosphere estimated by
employing the pn-QRPA theory were presented earlier in
Refs.~\citep{Nabi07} and~\citep{Muneeb13}, respectively. The authors
also compared their results with the previous calculations and
available experimental data. In present work, GT strength
distributions, gamma-heating as well as neutrino-cooling rates in
lepton emission and capture directions have been estimated for a
chain of twenty two vanadium isotopes ($^{43-64}$V). For a direct
comparison of our computed weak rates with the previously determined
IPM and LSSM rates, the same domain of temperature, density and grid
points were used.

In the coming section, the formalism of the pn-QRPA theory is
illustrated briefly. The results on GT strength distribution,
cooling rates and heating rates, as a result of weak-interaction
processes on vanadium isotopes in the astrophysical environment are
presented and discussed in Section 3. Section 4 concludes the main
findings of our work.

\section{Formalism}
\label{sec:formalism}

As stated earlier, we have made use of the pn-QRPA model for the
theoretical prediction of GT strength distributions and related
weak-interaction rates for vanadium (V) nuclides. The proton-neutron
residual interactions, in the pn-QRPA formalism occurr in two
different channels, that is particle-particle and particle-hole
interaction. The Hamiltonian ($H$) which has been used in our model
is given as
\begin{equation}
H = H^{SP} + V^{Pairing} + V_{GT}^{PP} + V_{GT}^{PH} .
\end{equation}
where $H^{SP}$ depicts the single-particle ($SP$) hamiltonian, The
pairing interaction is given by $V^{Pairing}$. $V_{GT}^{PP}$ and
$V_{GT}^{PH}$ represent the particle-particle ($PP$) and the
particle-hole ($PH$) GT forces, respectively. The energies and wave
functions of the single particle were estimated by applying the
Nilsson model~\citep{Nil55}. This model also takes into
consideration the nuclear deformation ($\beta_{2}$). BCS
approximation was employed to treat the pairing force. Two
interaction constants, namely $\kappa$ and $\chi$ were used to
characterize the $PP$ and $PH$ interactions, respectively. The
values of these constants were adjusted such that the pn-QRPA
estimated values of $\beta$-decay half-lives of vanadium isotopes
are compatible to the experimental values reported in
Ref.~\citep{Aud17} and also the Ikeda sum rule~\citep{Isr63} was
obeyed.
The commonly used values of pairing gaps
\begin{equation}
\Delta _{n} = \Delta _{p} = 12/\sqrt{A}\;(MeV).
\end{equation}
were adopted in this paper. Ref.~\citep{Nil55} was used to choose
the Nilsson potential parameters. The following formula was used to
determine the nuclear deformation values
\begin{equation}
\beta_{2} = \frac{125(Q_2)}{1.44(Z)(A^{2/3})} \label{Eq:deform}~,
\end{equation}
$Q_2$ in this relation represents the electric quadrupole moment
whose values were selected from Ref.~\citep{Mol95}. The reaction
Q-values were used from latest data from mass compilation of
Ref.~\citep{Aud17}.

In pn-QRPA model, the charge-exchange transformations are specified
in terms of phonon creation operators, where the pn-QRPA phonons are
specified as
\begin{equation}
A_{\omega}^{+}(\mu)=\sum_{pn}(X^{pn}_{\omega}(\mu)a_{p}^{+}a_{\bar{n}}^{+}-Y_{\omega}^{pn}(\mu)a_{n}
a_{\bar{p}}). \label{Eq:phonon}
\end{equation}
The summation in Eq.~\ref{Eq:phonon} was carried on all the
proton-neutron pairs, for $\mu$ = (\textit{m$_{p}$ - m$_{n}$}) = $0,
\pm 1$, where \textit{m$_{n}$}(\textit{m$_{p}$}) designates the
3$^{rd}$ component of angular momentum of neutron(proton). The
operator for the creation of a quasi-particle state of
neutron(proton) is represented by \textit{a$^{+}_{n(p)}$}. The QRPA
vacuum, designated as $|QRPA\rangle$, can be determined from the
annihilation condition A$_{\omega}(\mu)|QRPA\rangle$ = 0, and is
considered as the ground state of the model. The renowned RPA matrix
equation was solved to obtain the phonon operator
A$_{\omega}^{+}(\mu)$, with $\omega$ representing the excitation
energy and amplitudes (\textit{X$_{\omega}, Y_{\omega}$}). For a
complete solution of RPA matrix equation, reader is referred to
Refs.~\citep{Hir91, Mut89}.

The neutrinos and anti-neutrinos in stellar atmosphere are mainly
created via four specific types of weak-interaction mediated
reactions: by positron and electron emissions, and because of
continuum captures of electron and positron. As stated earlier,
these (anti)neutrinos during presupernova stages can escape from the
stellar core taking away energy and thus the core cools down. The
cooling rates due to (anti)neutrino emission are calculated as
\begin{eqnarray}
\lambda ^{^{\nu(\bar{\nu})}} _{nm} &=& \left(\frac{\ln 2}{D} \right)
[f_{nm}^{\nu} (\rho, T, E_{f})][B(F)_{nm} \nonumber \\
&+&(g_{A}/ g_{V})^{2} B(GT)_{nm}].\label{Eq:Nu}
\end{eqnarray}
The value of constant D in Eq.~\ref{Eq:Nu} was chosen as 6143s
\citep{Har09} and $g_{A}/g_{V}$ value was set equal to -1.2694
\citep{Nak10}. B(GT)$_{nm}$ and B(F)$_{nm}$ represent the sum of
reduced transition probabilities due to GT and Fermi interactions,
respectively and are determined as
\begin{eqnarray}
B(F)_{nm} &=& \frac{1}{2J_{n} + 1} |\langle m ||\sum_{k} \tau^{k}_{\pm} || n \rangle |^{2},\\
B(GT)_{nm} &=& \frac{1}{2J_{n} + 1} |\langle m ||\sum_{k} \tau^{k}_{\pm} \vec{\sigma}^{k} || n \rangle
|^{2},
\end{eqnarray}
where $\overrightarrow{\sigma}(k)$ stand for spin operator and
$\tau_{\pm }^{k}$ is iso-spin operator. For details on the
constitution of the excited levels of residual and parent nuclei and
calculation of nuclear matrix element within the structure of
pn-QRPA theory, reader is referred to Ref.~\citep{Nabi99}. The
($f_{nm}^{\nu}$) in Eq.~\ref{Eq:Nu} represent the phase space
integrals, which are functions of Fermi energy of electrons
($E_{f}$), stellar density and temperature. Explicit expressions of
$f_{nm}^{\nu}$ (calculated in natural units, $c = \hbar = m_{e} =
1$) for the decay channels, where for the positron (electron)
emission, lower (upper) sign is to be used, are given by
\begin{equation}
f_{nm}^{\nu} =\int _{1}^{\varepsilon_{m} } \varepsilon
(\varepsilon^{2} -1)^{1/2}(\varepsilon_{m} -\varepsilon)^{3}F(\pm
Z,\varepsilon)(1- G_{\mp })d\varepsilon. \label{phdecay}
\end{equation}
and for the lepton captures, where for continuum electron (positron) capture, upper (lower) sign is to be used, are
\begin{equation}
f_{nm}^{\nu} = \int _{\varepsilon_{l} }^{\infty }\varepsilon
(\varepsilon^{2} -1)^{1/2}(\varepsilon_{m} +\varepsilon)^{3} F(\pm
Z, \varepsilon) G_{\mp }d\varepsilon. \label{phcapture}
\end{equation}
In these equations, $\varepsilon$ represents the total (K.E + rest)
energy of the positron or electron.  $\varepsilon_{m}$ is the total
beta decay energy and $\varepsilon_{l}$ is the total threshold
energy for lepton capture. The Fermi functions, F($\pm
Z$,$\varepsilon$) were determined by employing the same procedure as
adopted by Ref.~\citep{Gove71}. G$_{+}$ (G$_{-}$) represent the
Fermi Dirac distribution functions for positron (electron) and are
given by
\begin{equation}
G_{+} = \frac{1}{e^{(E_{f} +E +2)/kT}+1}, \label{Gp}
\end{equation}
\begin{equation}
G_{-} = \frac{1}{e^{(E - E_{f})/kT}+1},
\label{Gm}
\end{equation}
where $k$ is Boltzman constant and E = ($\varepsilon$ - 1) is the
K.E. of the electron.

The total (anti)neutrino cooling rates per unit time to each nuclide
is calculated as
\begin{equation}
\lambda^{\nu(\bar{\nu})} =\sum _{nm}P_{n} \lambda
_{nm}^{\nu(\bar{\nu})}, \label{rate}
\end{equation}
where $P_{n}$ represents the occupation possibility of parent
excited levels and  obeys the Boltzmann statistical distribution.
The summation in Eq.~(\ref{rate}), was executed over whole set of
initial and final levels until an acceptable convergence was gained
in the estimation of rates.

Following relation was used to calculate the total gamma ray heating rate
\begin{equation}
\lambda_{\gamma} =\sum _{nm}P_{n} \lambda _{nm} E_{m},
\end{equation}
where $P_{n}$ is the occupation probability as described above.
$\lambda_{nm}$ represents the overall sum of the positron emission
and electron capture rates $\lambda_{nm} = \lambda_{nm}^{pe} +
\lambda_{nm}^{ec}$, or the sum total of electron emission and
positron capture rates, $\lambda_{nm} = \lambda_{nm}^{ee} +
\lambda_{nm}^{pc}$. $E_{m}$ depicts the energy of the residual
nucleus's excited levels.

\section{Results and discussions}
\label{sec:results}

The GT strength distributions, neutrino cooling rates
($\lambda^{\nu}$) and gamma heating rates ($\lambda_{\gamma}$) for
twenty two vanadium isotopes ($^{43-64}V$) have been calculated in
both lepton capture and beta-decay directions. $\beta$-decay
half-lives of these nuclides under terrestrial conditions were also
estimated using the pn-QRPA theory. As per previous studies, the GT
strengths calculated using different nuclear models are generally
higher than that measured experimentally~\citep{Vet89, Gaard83,
Roe93}. The calculated GT strengths are then renormalized by
different models by applying some fix value of quenching factor
($f_{q}$). In RPA calculations, the GT strengths are usually
quenched by a factor of 0.6~\citep{Vet89, Roe93}. The same value of
$f_{q}$ has been used in our present calculation. In renormalized
form, the Ikeda Sum rule (Re-ISR) in our model becomes:
\begin{equation*}
\text{Re-ISR} = \sum B(GT)_{-} - \sum B(GT)_{+}\cong
3f_{q}^{2}(N-Z). \label{Eqt. ISR}
\end{equation*}
Table~\ref{tab-ISR} shows a comparison of our calculated values of
Re-ISR with the theoretically predicted values. As it can be seen,
our model satisfied well the Ikeda sum rule. Table~\ref{tab-ISR}
also describes that the pn-QRPA estimated half-lives are in decent
accordance with the experimentally measured values taken
from~\citep{Aud17}. Note that $^{51}$V is stable and $^{50}$V is
nearly stable (T$_{1/2} = 1.5 \times 10^{17}$ years) isotope. The
nuclear deformation values calculated using Eq.~\ref{Eq:deform} for
our chain of isotopes are also given in the last column of
Table~\ref{tab-ISR}.

\begin{figure}
\begin{center}
\includegraphics[width=0.7\textwidth]{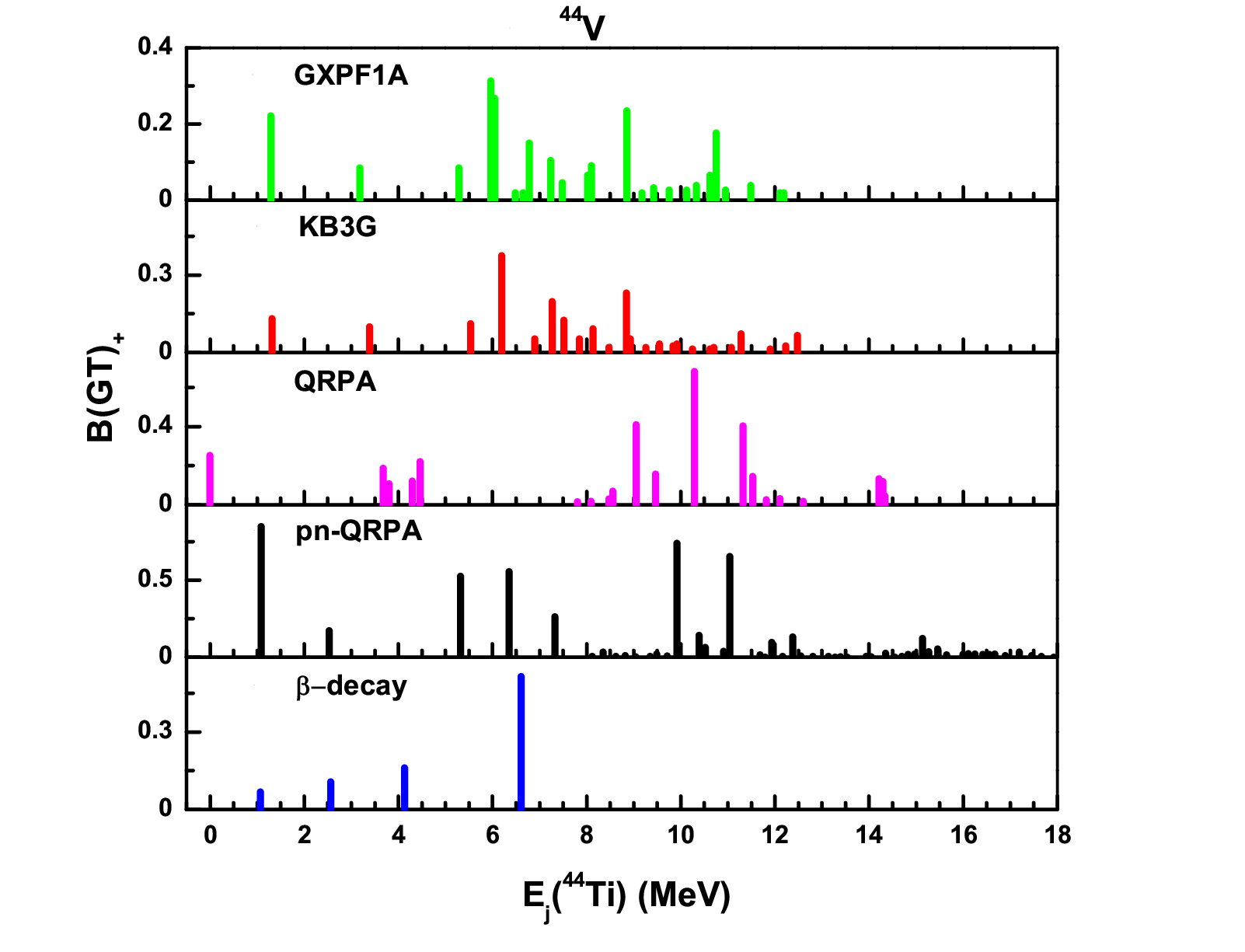}
\end{center}
\caption{Comparison of pn-QRPA calculated B(GT)$_{+}$ strength
distributions in $^{44}$V with experimental measurement and other
theoretical models (data taken from Ref.~\citep{Hitt16}).
GXPF1A~\citep{Honma05} and KB3G~\citep{Pove01} show shell model
calculations in the full pf-shell space, QRPA shows calculation done
using the formalism of Krumlinde and M\"{o}ller and $\beta$-decay
shows experimental data from~\citep{Hagberg97}.} \label{44V}
\end{figure}

\begin{figure}
\begin{center}
\includegraphics[width=0.7\textwidth]{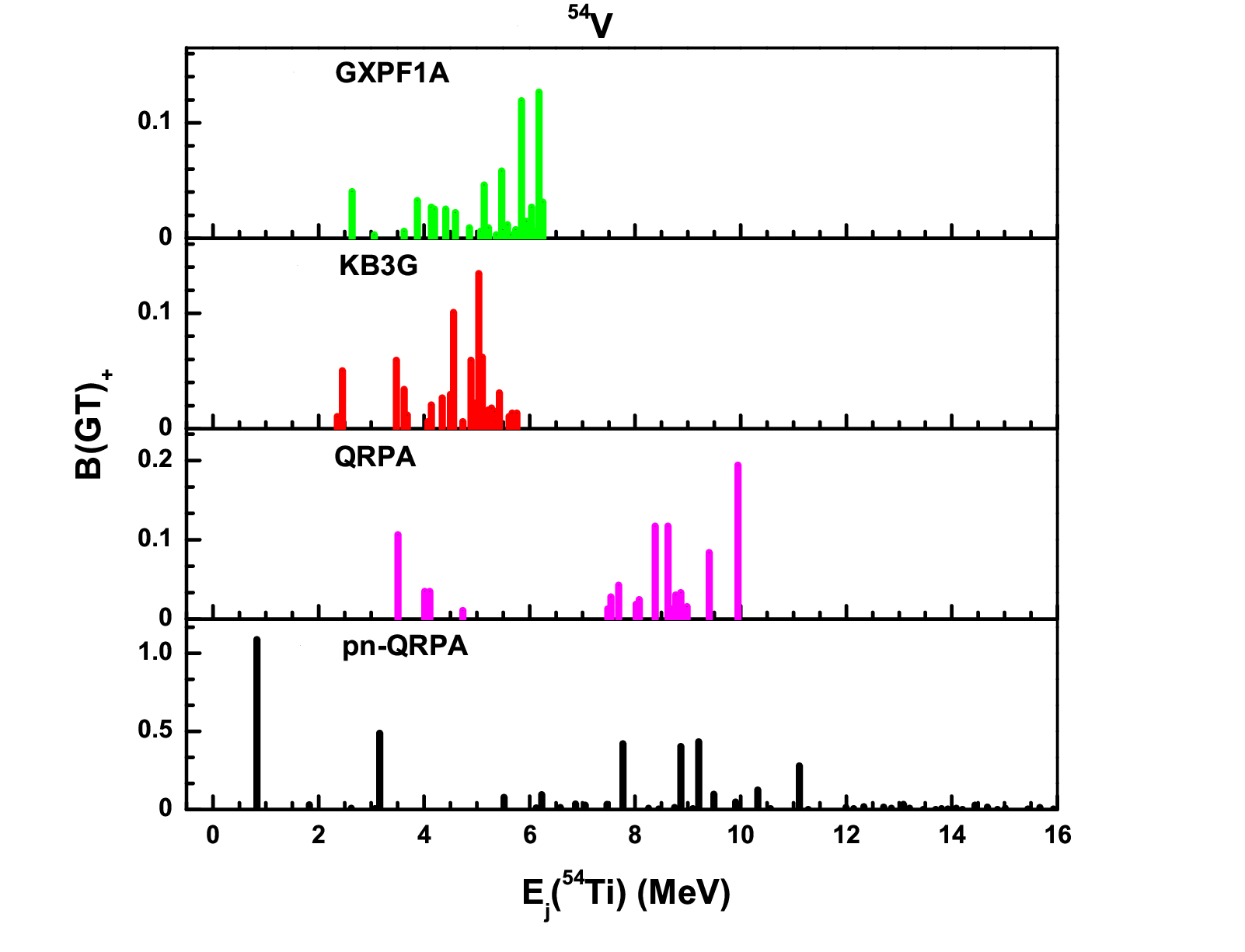}
\end{center}
\caption{Comparison of pn-QRPA calculated B(GT)$_{+}$ strength
distributions in $^{54}$V with other theoretical models. For details
see caption of Figure~\ref{44V}.} \label{54V}
\end{figure}

\begin{figure}
\begin{center}
\includegraphics[width=0.7\textwidth]{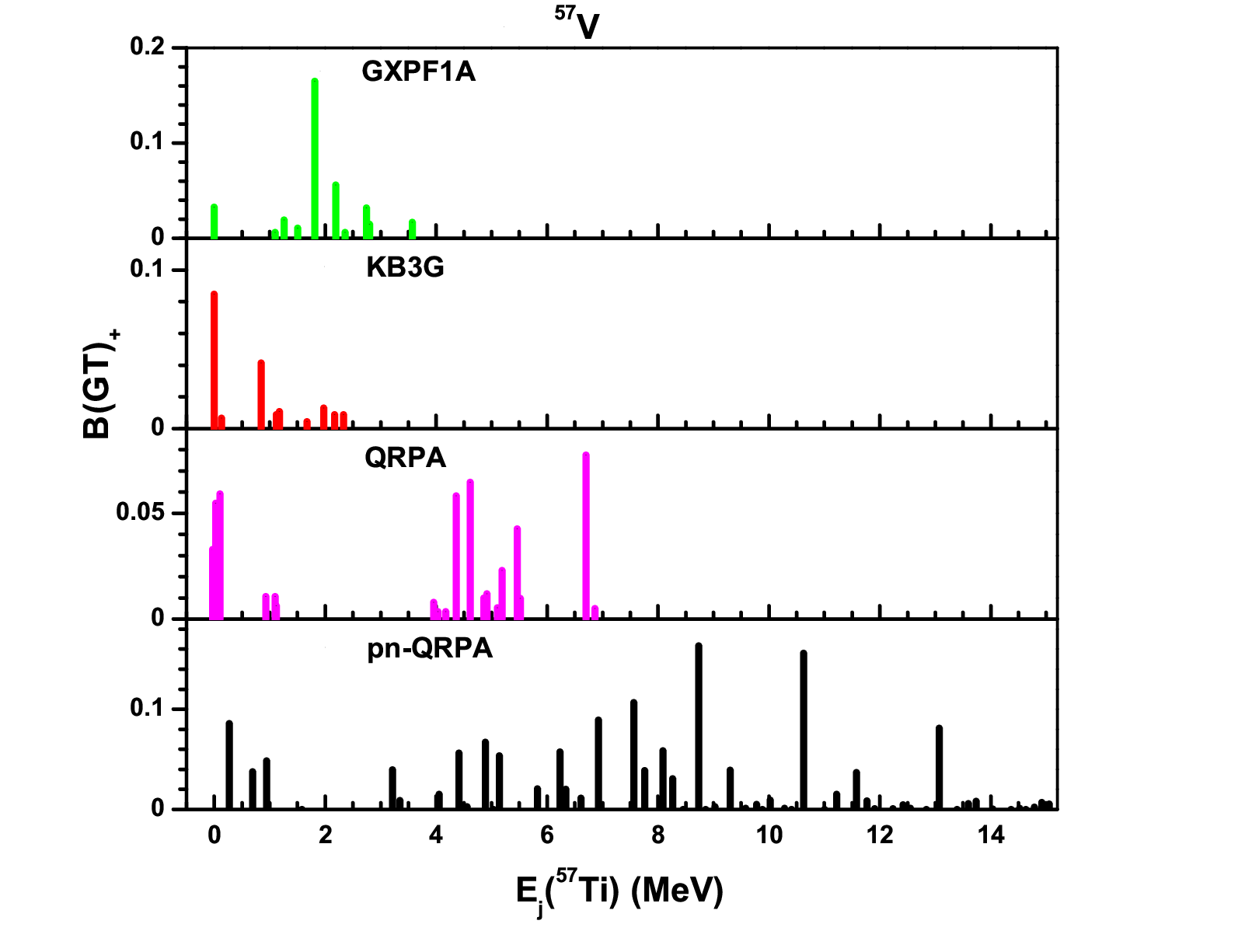}
\end{center}
\caption{Same as in Figure~\ref{54V}, but for B(GT)$_{+}$ strength
distributions in $^{57}$V.} \label{57V}
\end{figure}

\begin{figure}
\begin{center}
\includegraphics[width=0.7\textwidth]{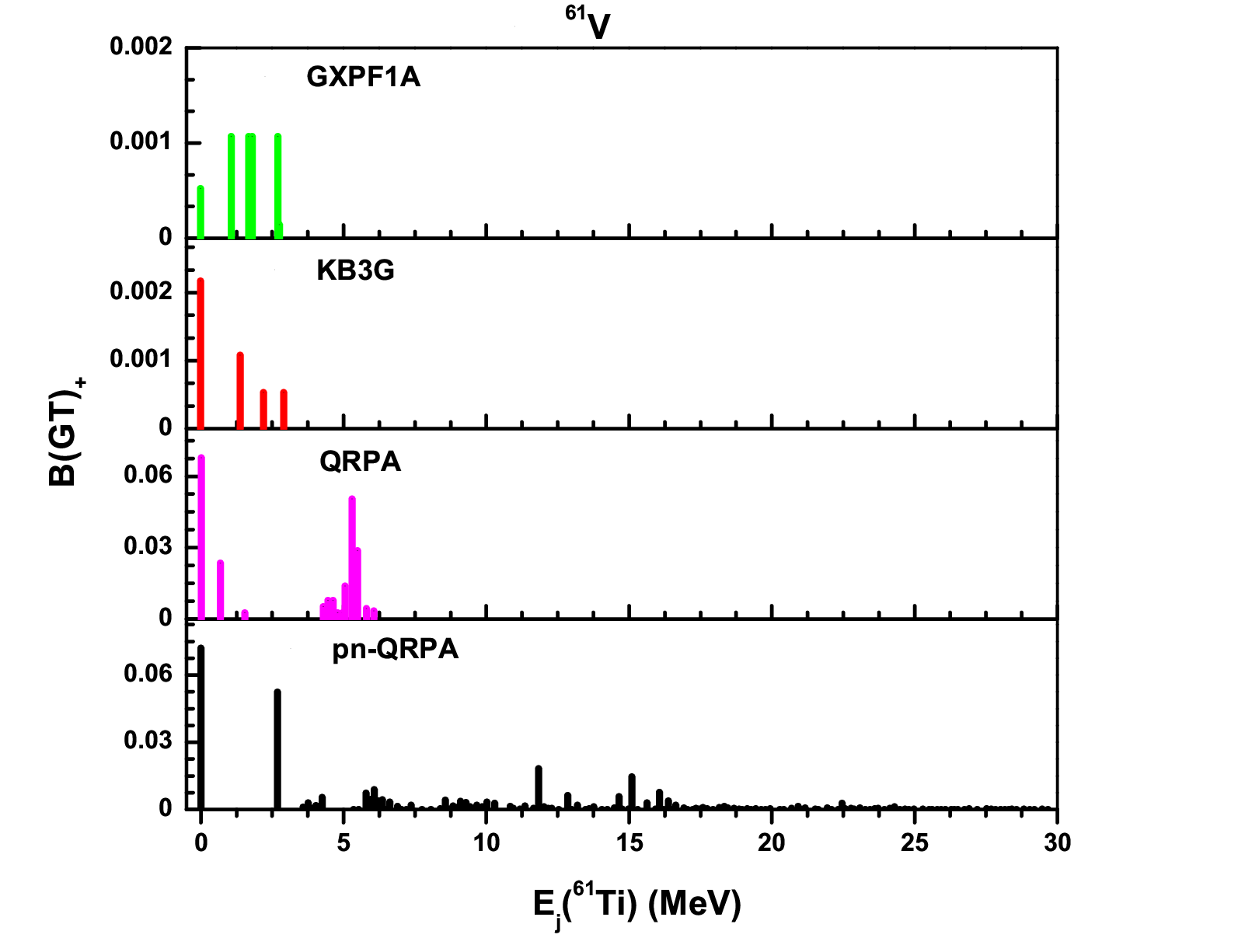}
\end{center}
\caption{Same as in Figure~\ref{54V}, but for B(GT)$_{+}$ strength
distributions in $^{61}$V.} \label{61V}
\end{figure}

In Figures.~\ref{44V}-\ref{61V}, our deduced B(GT)$_{+}$ strength
distributions for some of the selected vanadium nuclide as a
function of residual nucleus excitation energies in electron capture
direction are compared with the corresponding results
from~\citep{Hitt16}. They have presented B(GT) strengths computed
with shell model determined using GXPF1a~\citep{Honma05} and
KB3G~\citep{Pove01} fp-shell effective interactions, using QRPA
model in the formalism of Krumlinde and M\"{o}ller~\citep{Krum84}
and from experimental measurement (wherever available). One notes
that the GT strength is well fragmented in daughter states. For the
case of $^{44}$V the pn-QRPA model successfully reproduces the
measured peak around 6.5 MeV. For the case of $^{54,57,61}$V the
QRPA models calculate high-lying GT transitions as compared to the
shell model calculations.

Neutrino and anti-neutrino cooling rates as a result of
weak-interaction processes on $^{43-64}$V isotopes for a broad
extent of stellar densities and temperatures have been computed
using the pn-QRPA model. In Tables~\ref{Nu-V43-46}-\ref{Nu-V62-64},
the cooling rates have been shown at 4 selected values of stellar
densities ($\log \rho$Y$_{e}$ = 2, 5, 8, and 11\;g/cm$^{3}$), for
temperatures (T$_{9}$ given in units of $10^{9}$\;K) varying from
0.01 to 30. For every isotope of vanadium, there are two columns in
the tables, where the $\lambda^{\nu}$ ($\lambda^{\bar{\nu}}$) in the
first (second) column refers to the total neutrino (anti-neutrino)
cooling rates because of electron capture and positron emission
(positron capture and electron emission) and are stated in units of
MeV\;s$^{-1}$. It is to be noted that, 1.00E-100 in tables indicates
that the estimated rate is less than $1.00 \times 10^{-100}$. These
tables show that, neutrino and anti-neutrino cooling rates due to
weak-interaction processes on vanadium isotopes increase with the
the increase of core temperature. This is related to an increment in
the corresponding lepton capture and emission rates on these
isotopes with the rising temperature. At higher stellar densities,
where there is an increase (decrease) in Fermi energy (phase space
factor), beta-decay rates decrease and correspondingly reduction in
neutrino cooling rates is observed.

From Tables~\ref{Nu-V43-46} and~\ref{Nu-V47-50}, it can be seen
that, in general, neutrino cooling rates on $^{43-51}$V isotopes
are bigger than the corresponding anti-neutrino cooling rates by several orders of magnitude. For some of the isotopes
($^{48-51}$V), at T$_{9} = 30$, the anti-neutrino cooling rates try
to match with the neutrino cooling rates in the low and medium
density regions.  For vanadium isotopes with $A \ge 52$ (see
Tables~\ref{Nu-V51-54}-\ref{Nu-V62-64}), at lower temperatures, for
densities from $\log \rho$Y$_{e}$ = 2 -- 8\;g/cm$^{3}$, anti-neutrino cooling rates
dominate by several orders of magnitude. At higher temperature
(T$_{9}$ = 30), neutrino and anti-neutrino cooling rates for
$^{52-54}$V isotopes become comparable. However, as we move towards
more neutron rich isotopes, differences between the two rates
increase and the anti-neutrino cooling rates dominate by 1-2 orders
of magnitude. In the high density region ($\log \rho$Y$_{e}$ = 11),
the neutrino cooling rates again prevail the anti-neutrino cooling
rates.

\begin{figure}
\begin{center}
\includegraphics[width=0.8\textwidth]{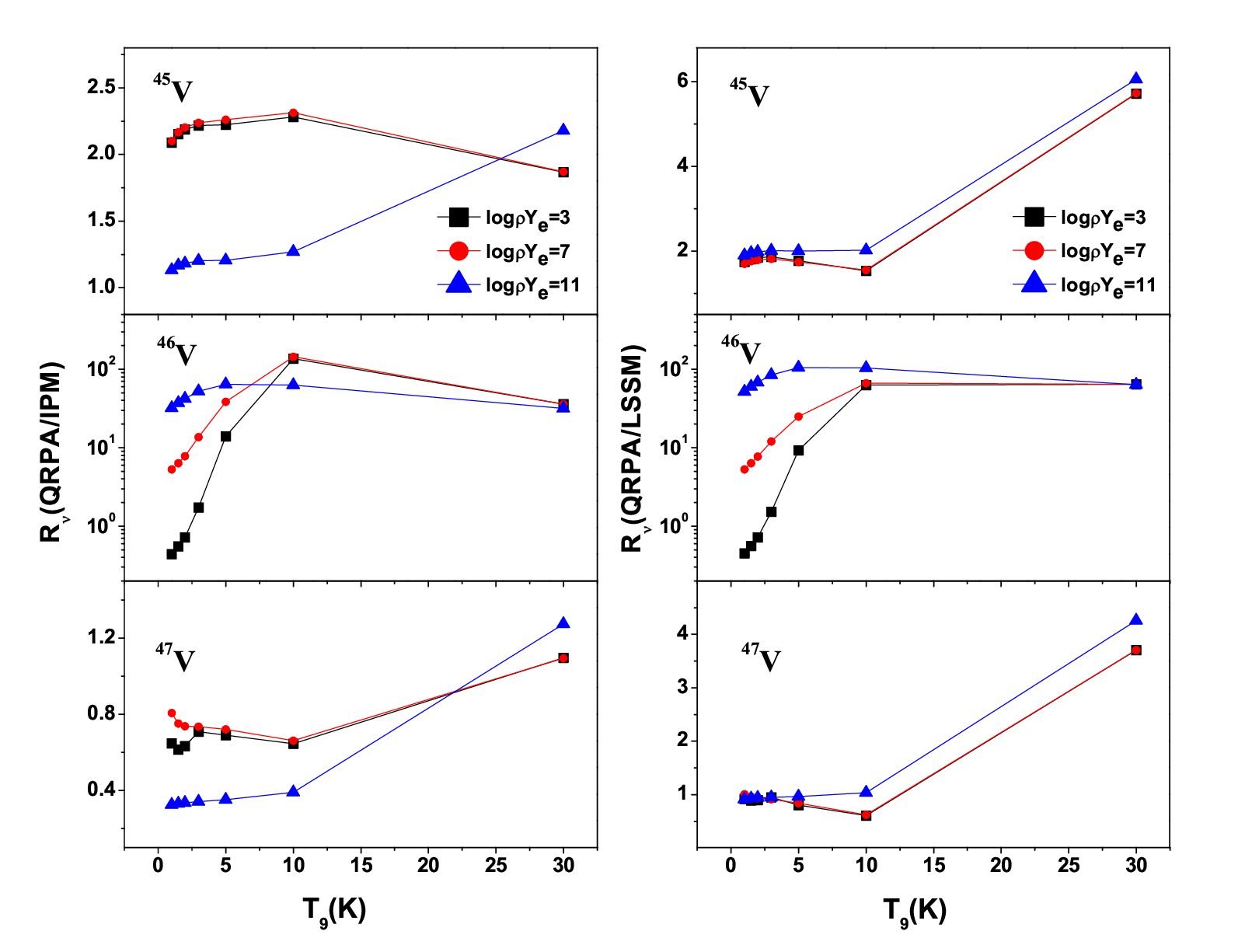}
\includegraphics[width=0.8\textwidth]{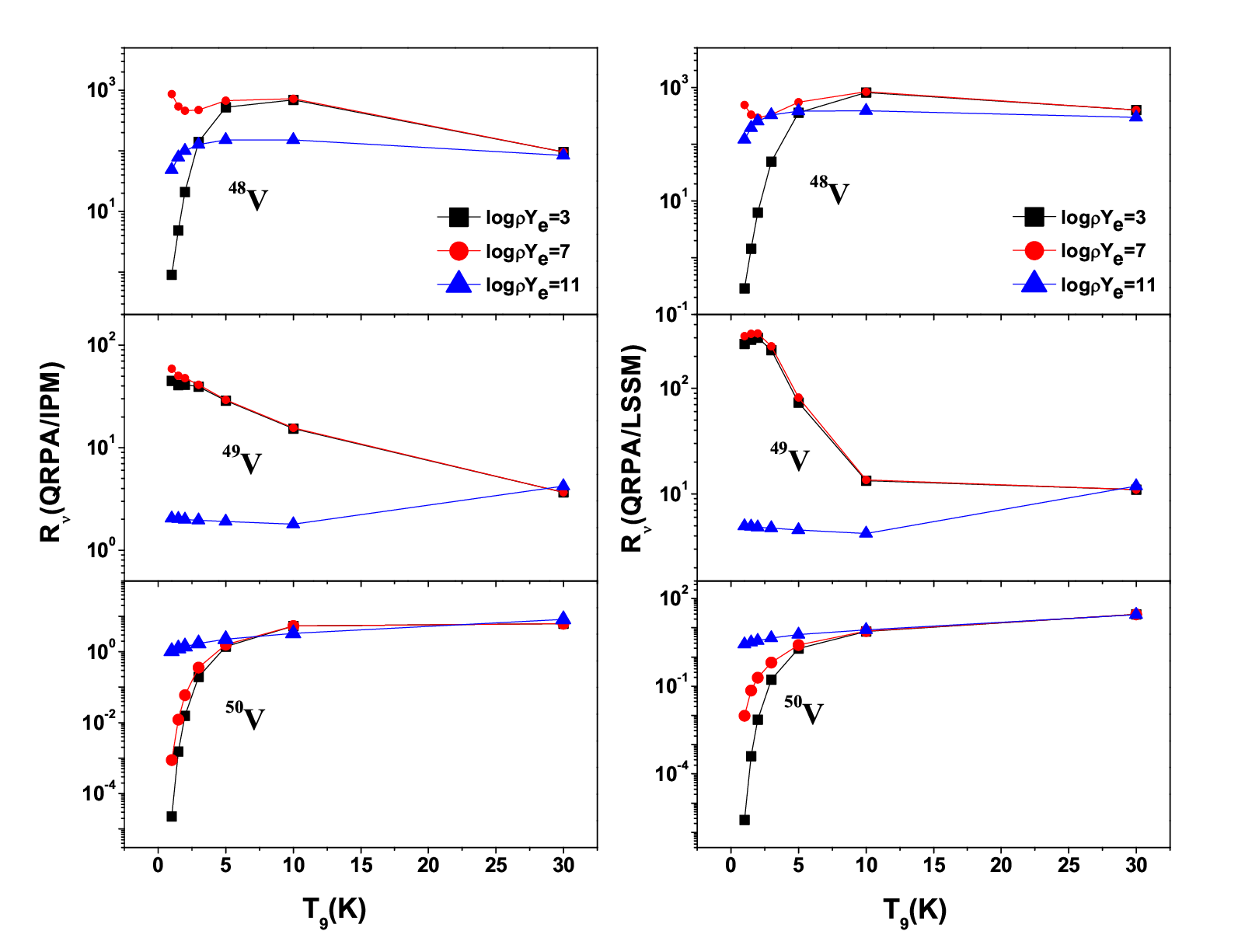}
\vspace{-0.5cm}\caption{The comparison of pn-QRPA calculated
	neutrino cooling rates due to$~^{45-50}$V with the previous
	calculations performed by IPM (on left) and LSSM (on right) at
	different selected densities as a function of stellar temperature.
	$\log\rho$Y$_{e}$ gives the $\log$ to base 10 of stellar density in
	units of g$\;$cm$^{-3}$.} \label{45-50-nu}
\end{center}
\end{figure}

\begin{figure}
\begin{center}
\includegraphics[height = 0.85\textwidth,width=1\textwidth]{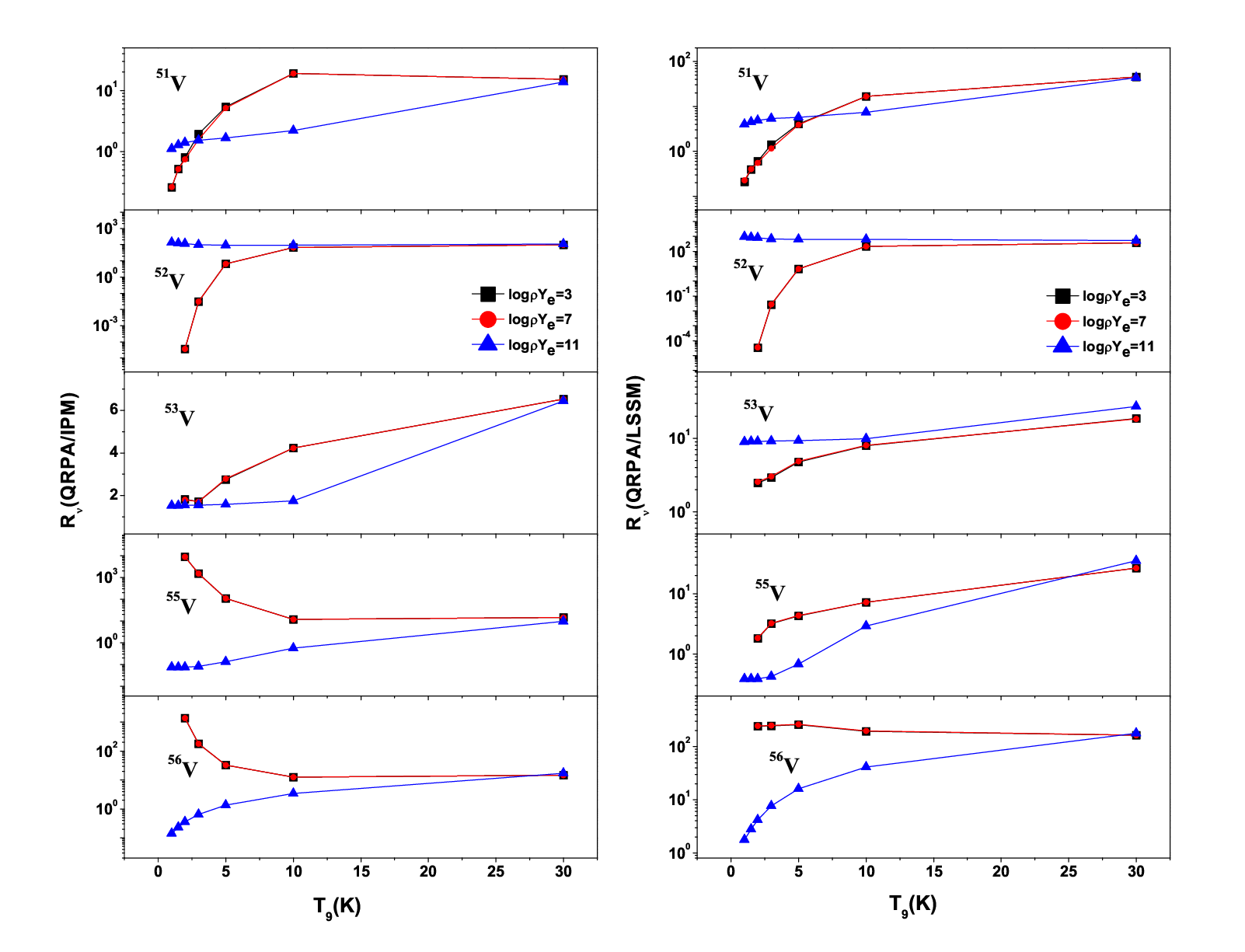}
\vspace{-0.5cm}\caption{Same as Figure~\ref{45-50-nu}, but
	for$~^{51-56}$V isotopes.} \label{51-56-nu}
\end{center}
\end{figure}

\begin{figure}
\begin{center}
\includegraphics[height = 0.85\textwidth,width=1\textwidth]{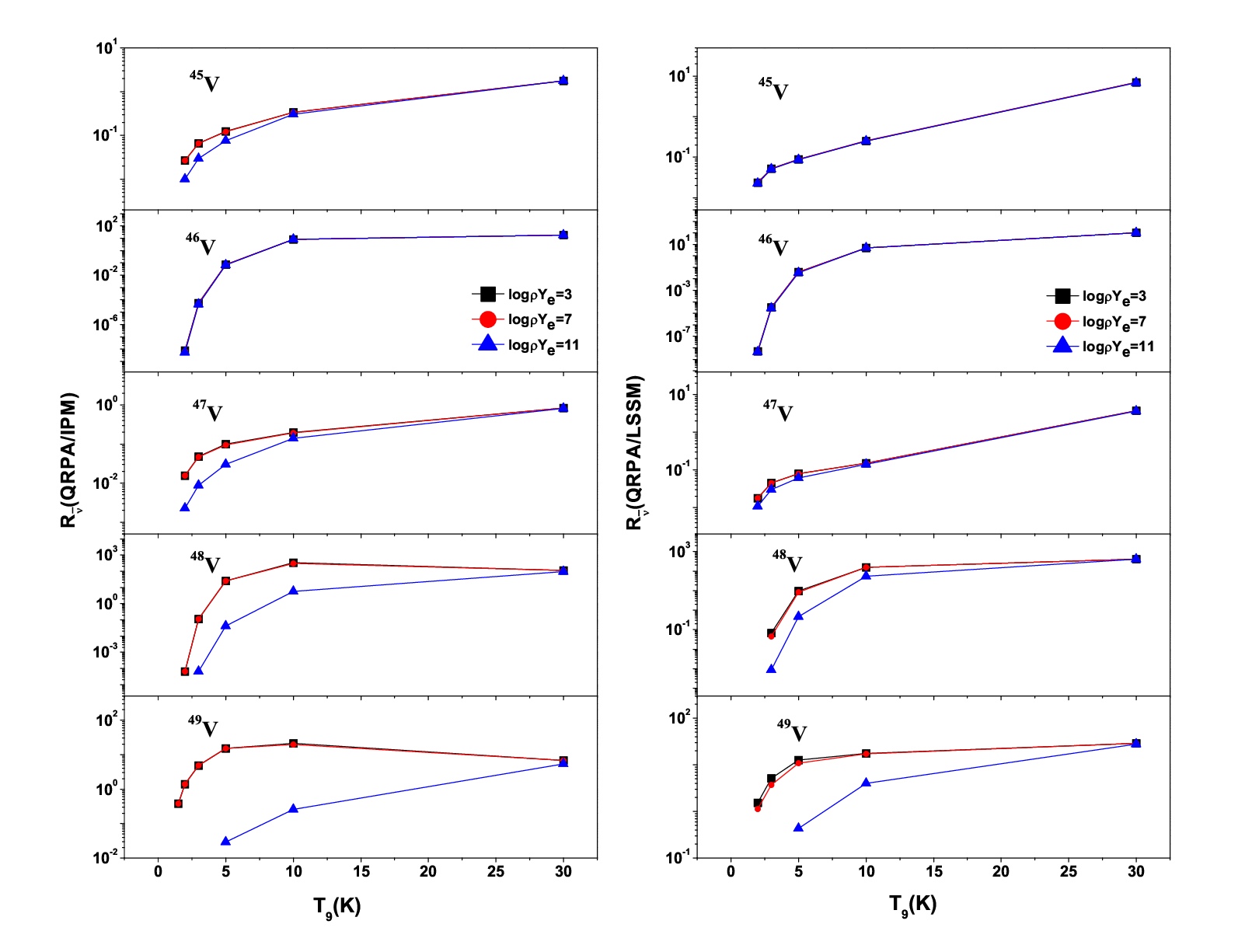}
\vspace{-0.5cm}\caption{The comparison of pn-QRPA calculated
	antineutrino cooling rates due to $^{45-49}$V isotopes with the
	previous calculations. Other details same as in
	Figure~\ref{45-50-nu}.} \label{45-49-nubar}
\end{center}
\end{figure}

\begin{figure}
\begin{center}
\includegraphics[height = 0.85\textwidth,width=1\textwidth]{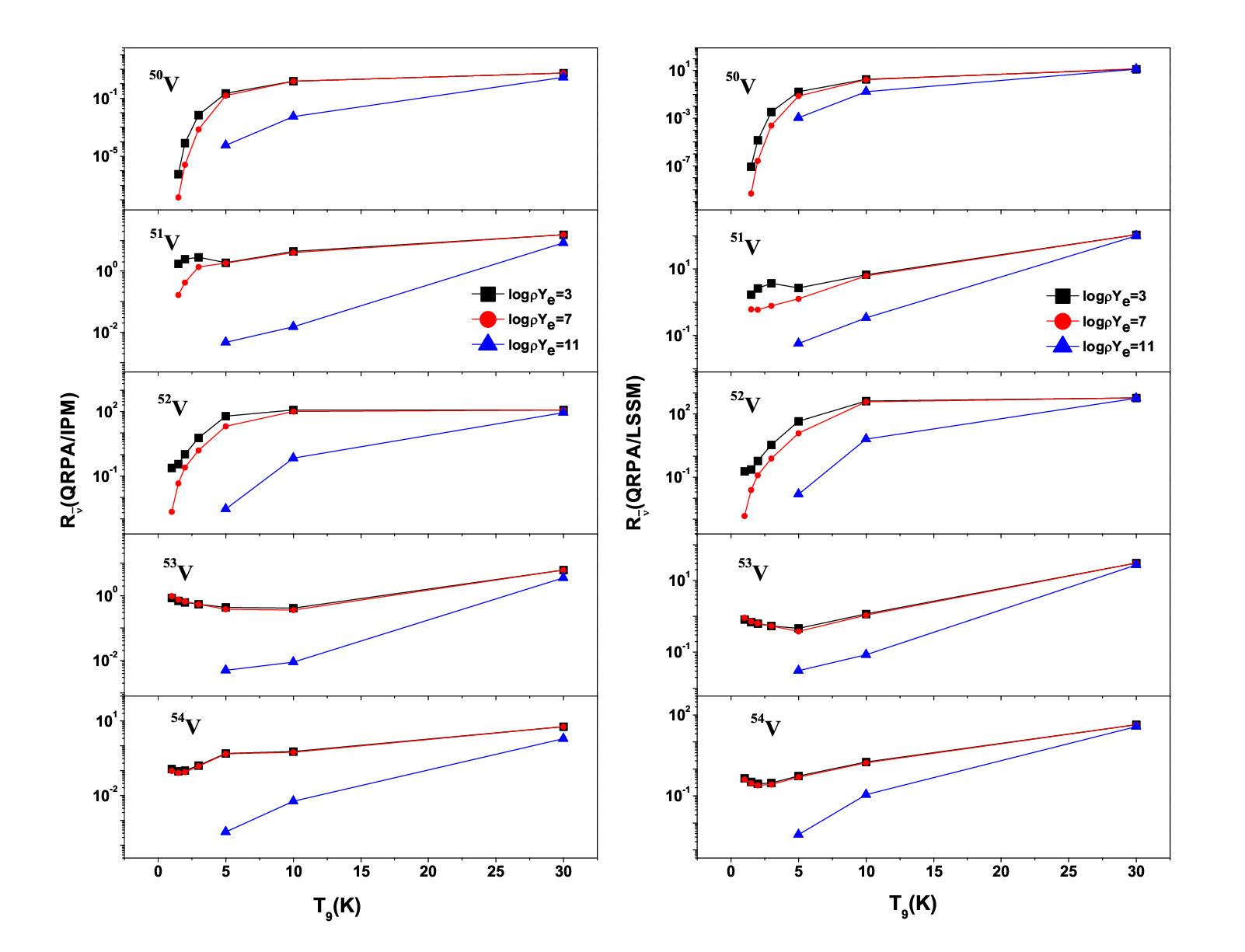}
\vspace{-0.5cm}\caption{Same as Figure~\ref{45-49-nubar}, but
	for$~^{50-54}$V isotopes.} \label{50-54-nubar}
\end{center}
\end{figure}

\begin{figure}
\begin{center}
\includegraphics[width=1\textwidth]{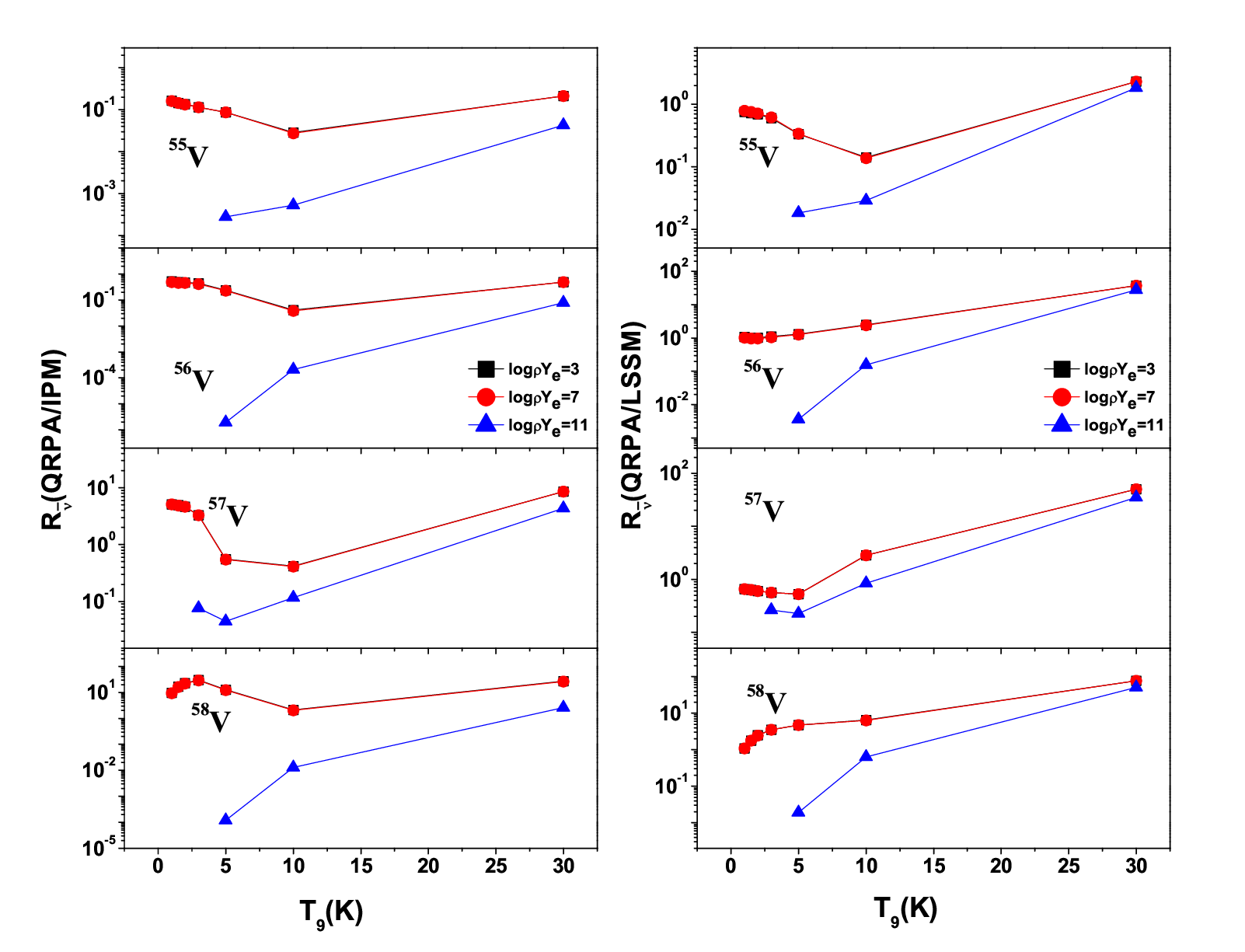}
\vspace{-0.5cm}\caption{Same as Figure~\ref{45-49-nubar}, but
	for$~^{55-58}$V isotopes.} \label{55-58-nubar}
\end{center}
\end{figure}

The neutrino and anti-neutrino cooling rates computed using the
deformed pn-QRPA theory have also been compared with the previously
calculated IPM and LSSM rates (wherever possible). The comparison
has been presented in graphical form in
Figures~\ref{45-50-nu}-\ref{55-58-nubar}. In these figures, the left
(right) panels show ratio of the current calculations to the IPM
(LSSM) calculations which are drawn as a function of stellar
temperature (T$_{9}$) at 3 values of densities ($\log \rho$Y$_{e}$ =
3, 7 and 11\;g/cm$^{3}$). The selected values of densities more or
less correspond to low-, medium- and high-density domains,
respectively. In general, it has been observed that at larger values
of temperature, where occupation probability of parent nucleus
excited levels increases, the pn-QRPA estimated neutrino and
anti-neutrino cooling rates enhance by up to 1-2 orders of magnitude
in contrast to previously reported IPM and LSSM computed rates.

Next we present the detailed analysis of comparison of neutrino
cooling rates shown in Figures~\ref{45-50-nu}-\ref{51-56-nu}.
Figure~\ref{45-50-nu} depicts that, cooling weak-rates due to
$^{45}$V and $^{47}$V calculated using three models are in
reasonable comparison (within a factor of $\sim$3) for most of the
density and temperature domain under study. Only, at large
temperatures (T$_{9}$ = 30), pn-QRPA computed rates for $^{45}$V
($^{47}$V) are factor 6 (4) larger than the corresponding LSSM
rates. In case of $^{46, 48, 49, 52, 53}$V, except at lower
temperatures and density (e.g., for $^{46}$V and $^{52}$V), our
estimated cooling weak rates are bigger as compared to previous
cooling rates calculated using IPM and LSSM models by up to 1-2
orders of magnitude. The difference in our rates and those
calculated with IPM and LSSM models may be ascribed to the
estimation of excited state GT transition strengths within different
models. The IPM and LSSM calculations applied the Brink hypothesis
to estimate the contributions to the GT transition strengths from
parent excited levels. This hypothesis asserts that GT strength
distributions for the exited state is similar to that for the ground
state, except that it is shifted by the state's excitation energy.
The pn-QRPA model, on the contrary, executes a state by state,
microscopic computation of GT strength distributions for all excited
levels of parent nucleus. The total weak rate is then evaluated by
taking sum over all parent and daughter levels.

For $^{50}$V and $^{51}$V, at low temperatures, in low and
medium-density domains, IPM and LSSM estimated rates are larger than
the pn-QRPA cooling weak rates. In the high-density region, the
mutual agreement between our rates and IPM rates improves, however
our rates exceed LSSM rates by up to a factor of 7-8. At (T$_{9}$ =
30) pn-QRPA cooling rates surpass both IPM and LSSM rates by factor
6-45. In case of $^{55, 56}$V, at lower densities and temperatures,
where the rates are smaller in magnitude, the larger differences
between the IPM and pn-QRPA rates are observed. Again, at extreme
temperatures, the cooling rates from the present calculations exceed
the results of IPM and LSSM by up to 1-2 orders of magnitude. In
addition to already stated difference for the estimation of GT
strength distributions between our and previous calculations (IPM,
and LSSM), there are some other possible reasons which may cause the
differences in their and our calculated rates. For example, the IPM
calculations suffered from the wrong placement of GT centriods, the
approximations used in nuclear matrix element calculations and
quenching of GT strengths. The LSSM calculations, however had issues
with the convergence as reported by authors in Ref.~\citep{Pruet03}.
Our model did not have any such issues and performs calculation in a
fully microscopic way.

Next, we turn to Figures~\ref{45-49-nubar}-\ref{55-58-nubar}, where
the comparison of anti-neutrino cooling rates as a result of weak
interactions on vanadium isotopes computed using different models is
presented. In general, at lower temperatures and high density, where
the estimated rates are small, larger differences are observed
between different calculations. In case of $^{45-47,50,54}$V, at
lower temperatures (T$_{9} \le 5$), the cooling rates calculated
using IPM and LSSM models are larger than our estimated rates by up
to 1-9 orders of magnitudes. With the rise of temperature, the
mutual comparison between the rates calculated using different
models gets better. At T$_{9}$ = 30, our computed rates get bigger
by factor of 2 to about 2 orders of magnitude. A roughly similar
kind of trend is observed in the comparison of cooling rates due to
$^{48, 49, 52}$V. For the next two cases, $^{51}$V and $^{53}$V, a
reasonable comparison is observed at lower temperatures and
densities. At higher temperatures, our rates exceed by up to an
order of magnitude. In case of $^{55, 56}$V, IPM reported rates
exceed the pn-QRPA rates by factor 2 to an order of magnitude. At
higher density, the enhancement is even up to 5 orders of magnitude.
Whereas, in $^{57, 58}$V, pn-QRPA cooling rates generally exceed
corresponding IPM rates. In case of $^{55-58}$V, at low and medium
densities, there is fair agreement between LSSM and pn-QRPA rates
for T$_{9} \le 5$. Again at higher temperatures, our rates surpass
the corresponding LSSM rates by up to one order of magnitude. The
differences in the rates from different models can again be
attributed to the reasons which are stated earlier.

The pn-QRPA computed gamma-heating rates on vanadium isotopes
($^{43-64}$V) are presented in
Tables~\ref{Gamma-V43-46}-\ref{Gamma-V62-64}. Unfortunately, to the
best of our knowledge, no previous calculations of gamma heating
rates on vanadium isotopes calculated using other models are
available in literature for comparison. In these tables, the selected
values of stellar densities and temperatures are given in the first
and second column, respectively.
$\lambda_{\gamma}^{\text{ec}+\text{pe}}$
($\lambda_{\gamma}^{\text{pc}+\text{ee}}$) represent gamma heating
rates due to electron capture and positron emission (positron
capture and electron emission). As stated earlier, beta decay and
electron capture rates increase with the increasing stellar
temperature and thus the associated gamma heating rates also
increase. Tables~\ref{Gamma-V43-46}-\ref{Gamma-V47-50} show that in
case of $^{43-50}$V gamma heating due to electron capture dominate
by orders of magnitude. For $^{52-64}$V (see
Tables~\ref{Gamma-V51-54}-\ref{Gamma-V62-64}), in the lower and
medium density domains, gamma heating rates prevail because of beta
decay and in the high density region because of electron capture. At
higher temperature T$_{9}$ = 30, for low and medium density regions,
gamma heating rates as a result of both electron capture and beta
decay on $^{51-56}$V are comparable.

\section{Conclusions}
\label{sec:conclusions}

Weak-interaction $\beta$-decay and electron capture processes on
iron-regime nuclei in the stellar core alter lepton to baryon
fraction (Y$_{e}$) during the late evolutionary stages of high-mass
stars. The temporal changes in the value of Y$_{e}$ in the core lead
to critical development in the collapse dynamics and to create an
explosion. At presupernova densities ($\sim 10^{11}$\;g/cm$^{3}$),
neutrinos and anti-neutrinos created due to these reactions freely
escape and cool the stellar interior and thereby core entropy is
reduced. The heat loss as a result of weak interaction neutrinos is
significantly reduced by the heating effect of gamma emission due to
electron captures and beta decay rates to the excited states. These heating and cooling
rates can affect the overall energy budget and entropy of the core
at any stage and control the onward progression. Therefore accurate
weak rates and the associated gamma heating as well as neutrino
cooling rates on astrophysically important nuclei is a crucial
nuclear input for modeling the presupernova and supernova stages of
stellar objects.

In this paper, we have employed deformed pn-QRPA theory which has a
good record to give reliable prediction of terrestrial $\beta$-decay
half lives, to estimate the stellar weak rates on a chain of twenty
two isotopes of vanadium ($^{43-64}$V). Our estimated values of
half-lives are in suitable comparison with the experimental data.
The GT strength distributions were calculated for all V isotopes in
both electron capture and $\beta$-decay directions which satisfied
well the model independent ISR. Neutrino cooling and gamma heating
rates have been calculated for a broad range of stellar density
($10-10^{11}\;$g/cm$^{3}$) and temperature ($0.01 \times 10^{9} - 30
\times 10^{9}$\;K). We have also compared our estimated cooling
rates with those of calculated using LSSM and IPM models (wherever
available). It is to be noted that, in the calculations of rates
performed using LSSM and IPM models, the Brink's hypothesis and back
resonances were used, which has not been employed by the current
pn-QRPA model. The IPM calculations also had other issues, like
misplacement of GT centroids and quenching of GT strengths. The LSSM
calculations, on the other hand experienced convergence issue (as
mentioned in Ref.~\citep{Pruet03}). Our model did not face such
problems and our calculations are totally microscopic.  From the
comparison of rates calculated with different models, it has been
observed that at higher temperature ($30 \times 10^{9}$\;K), our
computed cooling rates are enhanced by 1-2 orders of magnitude as
compared to the earlier calculations. Core-collapse simulators may
find our reported energy rates useful for modeling of presupernova
evolution of massive stars.

\section*{Acknowledgments}

J.-U. Nabi would like to acknowledge the support of the Higher Education Commission Pakistan through project numbers 5557/KPK/NRPU/R\&D/HEC/2016, 9-5(Ph-1-MG-7)/PAK-TURK/R\&D/HEC/2017 and Pakistan Science Foundation through project number PSF-TUBITAK/KP-GIKI (02).

\clearpage \onecolumn

\begin{table}
\centering
\caption{\small Comparison of pn-QRPA calculated $\beta$-decay
	half-lives (T$_{1/2}$) with measured data \citep{Aud17} (in units of
	second) and (re-normalized) calculated and theoretical Ikeda sum
	rule of $^{43-64}$V isotopes. The values of nuclear deformation
	$\beta_2$ used in our calculation are also given.}\label{tab-ISR}
\begin{tabular}{cccccc}
	&  &  &  &  &   \\
	\toprule
	Nuclei & T$_{1/2}$ (Cal)& T$_{1/2}$ (Exp) & Re-ISR$_{\text{Cal}}$ & Re-ISR$_{\text{Th}}$ &$\beta_{2}$ \\
	\midrule $^{43}$V & 8.14 $\times~10^{-2}$ & 7.93 $\times~10^{-2}$ &
	-3.29 & -3.24 & 0.011\tabularnewline $^{44}$V & 1.12
	$\times~10^{-1}$ & 1.11 $\times~10^{-1}$ & -2.21 & -2.14 &
	0.075\tabularnewline $^{45}$V & 6.66 $\times~10^{-1}$ & 5.47
	$\times~10^{-1}$ & -1.12 & -1.08 & 0.003\tabularnewline $^{46}$V &
	4.43 $\times~10^{-1}$ & 4.23 $\times~10^{-1}$ & -0.06 & 0.00 &
	0.019\tabularnewline $^{47}$V & 1.98 $\times~10^{3}$ & 1.96
	$\times~10^{3}$ & 1.02 & 1.08 & 0.205\tabularnewline $^{48}$V & 1.46
	$\times~10^{6}$ & 1.38 $\times~10^{6}$ & 2.20 & 2.14 &
	0.020\tabularnewline $^{49}$V & 2.91 $\times~10^{7}$ & 2.85
	$\times~10^{7}$& 3.31 & 3.24 & 0.161\tabularnewline $^{50}$V & - & -
	& 4.27 & 4.32 & 0.107\tabularnewline $^{51}$V & - & - &5.45 &5.40 &
	0.021 \tabularnewline $^{52}$V & 2.40 $\times~10^{2}$ & 2.24
	$\times~10^{2}$ & 6.41 & 6.48 & 0.053\tabularnewline $^{53}$V & 9.70
	$\times~10^{1}$ & 9.26 $\times~10^{1}$ & 7.52 & 7.56 &
	0.117\tabularnewline $^{54}$V & 5.11 $\times~10^{1}$ & 4.98
	$\times~10^{1}$ & 8.59 & 8.64 & 0.117\tabularnewline $^{55}$V & 6.65
	$\times~10^{0}$ & 6.54 $\times~10^{0}$ & 9.65 & 9.72 &
	0.172\tabularnewline $^{56}$V & 2.23 $\times~10^{-1}$ & 2.16
	$\times~10^{-1}$ & 10.86 & 10.80 & 0.173\tabularnewline $^{57}$V &
	3.55 $\times~10^{-1}$ & 3.50 $\times~10^{-1}$ & 11.81 & 11.88 &
	0.173\tabularnewline $^{58}$V & 1.94 $\times~10^{-1}$ & 1.91
	$\times~10^{-1}$ & 13.01 & 12.96 & 0.140\tabularnewline $^{59}$V &
	9.63 $\times~10^{-2}$ & 9.50 $\times~10^{-2}$ & 14.10 & 14.04 &
	0.140\tabularnewline $^{60}$V & 1.25 $\times~10^{-1}$ & 1.22
	$\times~10^{-1}$ & 15.27 & 15.12 & -0.135\tabularnewline $^{61}$V &
	4.96 $\times~10^{-2}$ & 4.82 $\times~10^{-2}$ & 16.14 & 16.20 &
	-0.104\tabularnewline $^{62}$V & 3.99 $\times~10^{-2}$ & 3.36
	$\times~10^{-2}$ & 17.19 & 17.28 & 0.011\tabularnewline $^{63}$V &
	1.99 $\times~10^{-2}$ & 1.96 $\times~10^{-2}$ & 18.31 & 18.36 &
	0.011\tabularnewline $^{64}$V & 1.54 $\times~10^{-2}$ & 1.50
	$\times~10^{-2}$ & 19.50 & 19.44 & -0.032\tabularnewline \bottomrule
\end{tabular}
\end{table}

\clearpage
\begin{table}[pt]
\caption{\small The pn-QRPA calculated neutrino and antineutrino
cooling rates due to weak rates on $^{43-46}$V at various selected
densities and temperatures in stellar environment. $\log\rho
\text{Y}_{e}$ has units of g/cm$^{3}$, where $\rho$ is the baryon
density and Y$_{e}$ is the ratio of the lepton number to the baryon
number. Temperature (T$_{9}$) is given in units of $10^{9}$\;K.
$\lambda^{\nu}$ ($\lambda^{\bar{\nu}}$) are the total neutrino
(antineutrino) energy loss rates  as a result of electron capture
and positron emission (positron capture and electron emission) in
units of MeV\;s$^{-1}$.}\label{Nu-V43-46} \centering {
\tiny
\begin{tabular}{cccccccccccc}
	& & & & & & & & &\\
	\toprule \multirow{2}{*}{$\log\rho$Y$_{e}$} &
	\multirow{2}{*}{T$_{9}$} & \multicolumn{2}{c}{$^{43}$V}&
	\multicolumn{2}{c}{$^{44}$V} & \multicolumn{2}{c}{$^{45}$V}&
	\multicolumn{2}{c}{$^{46}$V}\\
	\cmidrule{3-4}  \cmidrule{5-6}  \cmidrule{7-8} \cmidrule{9-10}& &
	\multicolumn{1}{c}{$\lambda^{\nu}$} &
	\multicolumn{1}{c}{$\lambda^{\bar{\nu}}$} &
	\multicolumn{1}{c}{$\lambda^{\nu}$} &
	\multicolumn{1}{c}{$\lambda^{\bar{\nu}}$} &
	\multicolumn{1}{c}{$\lambda^{\nu}$} &
	\multicolumn{1}{c}{$\lambda^{\bar{\nu}}$} &
	\multicolumn{1}{c}{$\lambda^{\nu}$} &
	\multicolumn{1}{c}{$\lambda^{\bar{\nu}}$} \\
	\midrule
	2.0 & 0.01 & 4.16E+01 & 1.00E-100 & 2.36E+01 & 1.00E-100 & 3.38E+00 & 1.00E-100 & 1.86E+00 & 1.00E-100\tabularnewline
	2.0 & 0.10 & 4.16E+01 & 1.00E-100 & 2.36E+01 & 1.00E-100 & 3.58E+00 & 1.00E-100 & 1.86E+00 & 1.00E-100\tabularnewline
	2.0 & 0.20 & 4.18E+01 & 1.00E-100 & 2.36E+01 & 1.00E-100 & 4.44E+00 & 1.00E-100 & 1.86E+00 & 1.00E-100\tabularnewline
	2.0 & 0.40 & 4.54E+01 & 1.00E-100 & 2.30E+01 & 1.00E-100 & 5.50E+00 & 1.00E-100 & 1.87E+00 & 1.00E-100\tabularnewline
	2.0 & 0.70 & 5.47E+01 & 1.00E-100 & 2.21E+01 & 1.00E-100 & 6.10E+00 & 2.34E-96 & 2.01E+00 & 5.65E-87\tabularnewline
\bottomrule
\end{tabular}}
\end{table}

\begin{table}[pt]
\caption{ Same as Table~\ref{Nu-V43-46}, but for $^{47-50}$V
isotopes.}\label{Nu-V47-50} \centering {\tiny
\begin{tabular}{cccccccccccc}
	& & & & & & & & &\\
	\toprule \multirow{2}{*}{$\log\rho$Y$_{e}$} &
	\multirow{2}{*}{T$_{9}$} & \multicolumn{2}{c}{$^{47}$V}&
	\multicolumn{2}{c}{$^{48}$V} & \multicolumn{2}{c}{$^{49}$V}&
	\multicolumn{2}{c}{$^{50}$V}\\
	\cmidrule{3-4}  \cmidrule{5-6}  \cmidrule{7-8} \cmidrule{9-10}& &
	\multicolumn{1}{c}{$\lambda^{\nu}$} &
	\multicolumn{1}{c}{$\lambda^{\bar{\nu}}$} &
	\multicolumn{1}{c}{$\lambda^{\nu}$} &
	\multicolumn{1}{c}{$\lambda^{\bar{\nu}}$} &
	\multicolumn{1}{c}{$\lambda^{\nu}$} &
	\multicolumn{1}{c}{$\lambda^{\bar{\nu}}$} &
	\multicolumn{1}{c}{$\lambda^{\nu}$} &
	\multicolumn{1}{c}{$\lambda^{\bar{\nu}}$} \\
	\midrule
	2.0 & 0.01 & 3.79E-04 & 1.00E-100 & 2.30E-06 & 1.00E-100 & 1.24E-07 & 1.00E-100 & 1.00E-100 & 1.00E-100\tabularnewline
	2.0 & 0.10 & 3.78E-04 & 1.00E-100 & 8.04E-07 & 1.00E-100 & 3.42E-08 & 1.00E-100 & 5.36E-73 & 1.00E-100\tabularnewline
	2.0 & 0.20 & 3.78E-04 & 1.00E-100 & 6.14E-07 & 1.00E-100 & 2.17E-08 & 1.35E-80 & 5.25E-42 & 2.67E-73\tabularnewline
	2.0 & 0.40 & 3.77E-04 & 1.00E-100 & 5.43E-07 & 1.08E-73 & 1.59E-08 & 5.42E-43 & 5.60E-26 & 5.05E-41\tabularnewline
	2.0 & 0.70 & 3.69E-04 & 3.13E-61 & 1.07E-06 & 2.10E-41 & 1.68E-08 & 2.88E-26 & 1.21E-18 & 7.80E-26\tabularnewline
\bottomrule
\end{tabular}}
\end{table}

\begin{table}[pt]
\caption{ Same as Table~\ref{Nu-V43-46}, but for $^{51-54}$V
isotopes.}\label{Nu-V51-54} \centering {\tiny
\begin{tabular}{cccccccccccc}
	& & & & & & & & &\\
	\toprule \multirow{2}{*}{$\log\rho$Y$_{e}$} &
	\multirow{2}{*}{T$_{9}$} & \multicolumn{2}{c}{$^{51}$V}&
	\multicolumn{2}{c}{$^{52}$V} & \multicolumn{2}{c}{$^{53}$V}&
	\multicolumn{2}{c}{$^{54}$V}\\
	\cmidrule{3-4}  \cmidrule{5-6}  \cmidrule{7-8} \cmidrule{9-10}& &
	\multicolumn{1}{c}{$\lambda^{\nu}$} &
	\multicolumn{1}{c}{$\lambda^{\bar{\nu}}$} &
	\multicolumn{1}{c}{$\lambda^{\nu}$} &
	\multicolumn{1}{c}{$\lambda^{\bar{\nu}}$} &
	\multicolumn{1}{c}{$\lambda^{\nu}$} &
	\multicolumn{1}{c}{$\lambda^{\bar{\nu}}$} &
	\multicolumn{1}{c}{$\lambda^{\nu}$} &
	\multicolumn{1}{c}{$\lambda^{\bar{\nu}}$} \\
	\midrule
	2.0 & 0.01 & 1.00E-100 & 1.00E-100 & 1.00E-100 & 9.33E-04 & 1.00E-100 & 1.24E-02 & 1.00E-100 & 2.62E-02\tabularnewline
	2.0 & 0.10 & 1.00E-100 & 5.87E-64 & 1.00E-100 & 9.35E-04 & 1.00E-100 & 1.24E-02 & 1.00E-100 & 2.62E-02\tabularnewline
	2.0 & 0.20 & 7.80E-76 & 8.28E-37 & 1.00E-100 & 9.35E-04 & 1.00E-100 & 1.24E-02 & 1.00E-100 & 2.62E-02\tabularnewline
	2.0 & 0.40 & 7.33E-44 & 6.92E-21 & 3.33E-76 & 9.35E-04 & 1.52E-73 & 1.23E-02 & 1.66E-92 & 2.62E-02\tabularnewline
	2.0 & 0.70 & 2.23E-29 & 3.65E-13 & 4.97E-46 & 9.23E-04 & 7.98E-46 & 1.21E-02 & 8.83E-56 & 2.68E-02\tabularnewline
\bottomrule
\end{tabular}}
\end{table}

\begin{table}[pt]
\caption{ Same as Table~\ref{Nu-V43-46}, but for $^{55-58}$V
isotopes.}\label{Nu-V55-58} \centering {\tiny
\begin{tabular}{cccccccccccc}
	& & & & & & & & &\\
	\toprule \multirow{2}{*}{$\log\rho$Y$_{e}$} &
	\multirow{2}{*}{T$_{9}$} & \multicolumn{2}{c}{$^{55}$V}&
	\multicolumn{2}{c}{$^{56}$V} & \multicolumn{2}{c}{$^{57}$V}&
	\multicolumn{2}{c}{$^{58}$V}\\
	\cmidrule{3-4}  \cmidrule{5-6}  \cmidrule{7-8} \cmidrule{9-10}& &
	\multicolumn{1}{c}{$\lambda^{\nu}$} &
	\multicolumn{1}{c}{$\lambda^{\bar{\nu}}$} &
	\multicolumn{1}{c}{$\lambda^{\nu}$} &
	\multicolumn{1}{c}{$\lambda^{\bar{\nu}}$} &
	\multicolumn{1}{c}{$\lambda^{\nu}$} &
	\multicolumn{1}{c}{$\lambda^{\bar{\nu}}$} &
	\multicolumn{1}{c}{$\lambda^{\nu}$} &
	\multicolumn{1}{c}{$\lambda^{\bar{\nu}}$} \\
	\midrule
	2.0 & 0.01 & 1.00E-100 & 3.27E-01 & 1.00E-100 & 9.08E+00 & 1.00E-100 & 8.05E+00 & 1.00E-100 & 1.54E+01\tabularnewline
	2.0 & 0.10 & 1.00E-100 & 2.78E-01 & 1.00E-100 & 9.06E+00 & 1.00E-100 & 6.81E+00 & 1.00E-100 & 1.54E+01\tabularnewline
	2.0 & 0.20 & 1.00E-100 & 2.51E-01 & 1.00E-100 & 8.95E+00 & 1.00E-100 & 6.41E+00 & 1.00E-100 & 1.54E+01\tabularnewline
	2.0 & 0.40 & 1.00E-100 & 2.36E-01 & 1.22E-95 & 8.47E+00 & 1.00E-100 & 6.21E+00 & 1.00E-100 & 1.55E+01\tabularnewline
	2.0 & 0.70 & 4.58E-63 & 2.28E-01 & 1.67E-57 & 8.00E+00 & 1.60E-83 & 6.10E+00 & 5.15E-75 & 1.77E+01\tabularnewline
\bottomrule
\end{tabular}}
\end{table}

\begin{table}[pt]
\caption{Same as Table~\ref{Nu-V43-46}, but for $^{59-61}$V
isotopes.}\label{Nu-V59-61} \centering {\tiny
\begin{tabular}{cccccccccc}
	& & & & & & & \\
	\toprule \multirow{2}{*}{$\log\rho$Y$_{e}$} &
	\multirow{2}{*}{T$_{9}$} & \multicolumn{2}{c}{$^{59}$V}&
	\multicolumn{2}{c}{$^{60}$V} & \multicolumn{2}{c}{$^{61}$V}\\
	\cmidrule{3-4}  \cmidrule{5-6}  \cmidrule{7-8} & &
	\multicolumn{1}{c}{$\lambda^{\nu}$} &
	\multicolumn{1}{c}{$\lambda^{\bar{\nu}}$} &
	\multicolumn{1}{c}{$\lambda^{\nu}$} &
	\multicolumn{1}{c}{$\lambda^{\bar{\nu}}$} &
	\multicolumn{1}{c}{$\lambda^{\nu}$} &
	\multicolumn{1}{c}{$\lambda^{\bar{\nu}}$} \\
	\midrule
	2.0 & 0.01 & 1.00E-100 & 3.40E+01 & 1.00E-100 & 2.79E+01 & 1.00E-100 & 6.71E+01\tabularnewline
	2.0 & 0.10 & 1.00E-100 & 3.40E+01 & 1.00E-100 & 2.79E+01 & 1.00E-100 & 6.71E+01\tabularnewline
	2.0 & 0.20 & 1.00E-100 & 3.38E+01 & 1.00E-100 & 2.79E+01 & 1.00E-100 & 6.71E+01\tabularnewline
	2.0 & 0.40 & 1.00E-100 & 3.25E+01 & 1.00E-100 & 2.79E+01 & 1.00E-100 & 6.71E+01\tabularnewline
	2.0 & 0.70 & 2.96E-96 & 3.09E+01 & 5.58E-85 & 2.82E+01 & 1.00E-100 & 6.76E+01\tabularnewline
\bottomrule
\end{tabular}}
\end{table}

\begin{table}[pt]
\caption{ Same as Table~\ref{Nu-V43-46}, but for $^{62-64}$V
isotopes.}\label{Nu-V62-64} \centering {\tiny
\begin{tabular}{cccccccccc}
	& & & & & & & \\
	\toprule \multirow{2}{*}{$\log\rho$Y$_{e}$} &
	\multirow{2}{*}{T$_{9}$} & \multicolumn{2}{c}{$^{62}$V}&
	\multicolumn{2}{c}{$^{63}$V} & \multicolumn{2}{c}{$^{64}$V}\\
	\cmidrule{3-4}  \cmidrule{5-6}  \cmidrule{7-8} & &
	\multicolumn{1}{c}{$\lambda^{\nu}$} &
	\multicolumn{1}{c}{$\lambda^{\bar{\nu}}$} &
	\multicolumn{1}{c}{$\lambda^{\nu}$} &
	\multicolumn{1}{c}{$\lambda^{\bar{\nu}}$} &
	\multicolumn{1}{c}{$\lambda^{\nu}$} &
	\multicolumn{1}{c}{$\lambda^{\bar{\nu}}$} \\
	\midrule
	2.0 & 0.01 & 1.00E-100 & 1.17E+02 & 1.00E-100 & 1.88E+02 & 1.00E-100 & 2.37E+02\tabularnewline
	2.0 & 0.10 & 1.00E-100 & 1.17E+02 & 1.00E-100 & 1.88E+02 & 1.00E-100 & 2.37E+02\tabularnewline
	2.0 & 0.20 & 1.00E-100 & 1.17E+02 & 1.00E-100 & 1.90E+02 & 1.00E-100 & 2.37E+02\tabularnewline
	2.0 & 0.40 & 1.00E-100 & 1.35E+02 & 1.00E-100 & 2.11E+02 & 1.00E-100 & 2.39E+02\tabularnewline
	2.0 & 0.70 & 3.40E-99 & 2.14E+02 & 1.00E-100 & 2.56E+02 & 1.00E-100 & 2.61E+02\tabularnewline
\bottomrule
\end{tabular}}
\end{table}

\begin{table}[pt]
\caption{\small The pn-QRPA calculated gamma heating rates due to
weak rates on $^{43-46}$V at various selected densities and
temperatures in stellar environment. $\log\rho \text{Y}_{e}$ has
units of g/cm$^{3}$, where $\rho$ is the baryon density and Y$_{e}$
is the ratio of the lepton number to the baryon number. Temperature
(T$_{9}$) is given in units of $10^{9}$\;K.
$\lambda_{\gamma}^{\text{ec}+\text{pe}}$
($\lambda_{\gamma}^{\text{pc}+\text{ee}}$) are the gamma heating
rates as a result of electron capture and positron emission
(positron capture and electron emission) in units of
MeV\;s$^{-1}$.}\label{Gamma-V43-46} \centering {\tiny
\begin{tabular}{cccccccccccc}
	& & & & & & & & &\\
	\toprule \multirow{2}{*}{$\log\rho$Y$_{e}$} &
	\multirow{2}{*}{T$_{9}$} & \multicolumn{2}{c}{$^{43}$V}&
	\multicolumn{2}{c}{$^{44}$V} & \multicolumn{2}{c}{$^{45}$V}&
	\multicolumn{2}{c}{$^{46}$V}\\
	\cmidrule{3-4}  \cmidrule{5-6}  \cmidrule{7-8} \cmidrule{9-10}& &
	\multicolumn{1}{c}{$\lambda_{\gamma}^{\text{ec}+\text{pe}}$} &
	\multicolumn{1}{c}{$\lambda_{\gamma}^{\text{pc}+\text{ee}}$} &
	\multicolumn{1}{c}{$\lambda_{\gamma}^{\text{ec}+\text{pe}}$} &
	\multicolumn{1}{c}{$\lambda_{\gamma}^{\text{pc}+\text{ee}}$} &
	\multicolumn{1}{c}{$\lambda_{\gamma}^{\text{ec}+\text{pe}}$} &
	\multicolumn{1}{c}{$\lambda_{\gamma}^{\text{pc}+\text{ee}}$} &
	\multicolumn{1}{c}{$\lambda_{\gamma}^{\text{ec}+\text{pe}}$} &
	\multicolumn{1}{c}{$\lambda_{\gamma}^{\text{pc}+\text{ee}}$} \\
	\midrule
	2.0 & 0.01 & 7.96E+00 & 1.00E-100 & 3.16E+01 & 1.00E-100 & 6.01E-03 & 1.00E-100 & 6.12E+00 & 1.00E-100\tabularnewline
	2.0 & 0.10 & 7.96E+00 & 1.00E-100 & 3.16E+01 & 1.00E-100 & 6.41E-03 & 1.00E-100 & 6.11E+00 & 1.00E-100\tabularnewline
	2.0 & 0.20 & 8.00E+00 & 1.00E-100 & 3.16E+01 & 1.00E-100 & 8.07E-03 & 1.00E-100 & 6.11E+00 & 1.00E-100\tabularnewline
	2.0 & 0.40 & 8.75E+00 & 1.00E-100 & 3.08E+01 & 1.00E-100 & 1.01E-02 & 1.00E-100 & 6.14E+00 & 1.00E-100\tabularnewline
	2.0 & 0.70 & 1.07E+01 & 1.00E-100 & 2.96E+01 & 1.00E-100 & 1.14E-02 & 1.00E-100 & 6.59E+00 & 2.04E-86\tabularnewline
\bottomrule
\end{tabular}}
\end{table}

\begin{table}[pt]
\caption{ Same as Table~\ref{Gamma-V43-46}, but for
$^{47-50}$V isotopes.}\label{Gamma-V47-50} \centering {\tiny
\begin{tabular}{cccccccccccc}
	& & & & & & & & &\\
	\toprule \multirow{2}{*}{$\log\rho$Y$_{e}$} &
	\multirow{2}{*}{T$_{9}$} & \multicolumn{2}{c}{$^{47}$V}&
	\multicolumn{2}{c}{$^{48}$V} & \multicolumn{2}{c}{$^{49}$V}&
	\multicolumn{2}{c}{$^{50}$V}\\
	\cmidrule{3-4}  \cmidrule{5-6}  \cmidrule{7-8} \cmidrule{9-10}& &
	\multicolumn{1}{c}{$\lambda_{\gamma}^{\text{ec}+\text{pe}}$} &
	\multicolumn{1}{c}{$\lambda_{\gamma}^{\text{pc}+\text{ee}}$} &
	\multicolumn{1}{c}{$\lambda_{\gamma}^{\text{ec}+\text{pe}}$} &
	\multicolumn{1}{c}{$\lambda_{\gamma}^{\text{pc}+\text{ee}}$} &
	\multicolumn{1}{c}{$\lambda_{\gamma}^{\text{ec}+\text{pe}}$} &
	\multicolumn{1}{c}{$\lambda_{\gamma}^{\text{pc}+\text{ee}}$} &
	\multicolumn{1}{c}{$\lambda_{\gamma}^{\text{ec}+\text{pe}}$} &
	\multicolumn{1}{c}{$\lambda_{\gamma}^{\text{pc}+\text{ee}}$} \\
	\midrule
	2.0 & 0.01 & 7.67E-06 & 1.00E-100 & 1.59E-05 & 1.00E-100 & 1.24E-08 & 1.00E-100 & 1.00E-100 & 1.00E-100\tabularnewline
	2.0 & 0.10 & 7.60E-06 & 1.00E-100 & 5.48E-06 & 1.00E-100 & 3.37E-09 & 1.00E-100 & 6.89E-71 & 1.00E-100\tabularnewline
	2.0 & 0.20 & 7.60E-06 & 1.00E-100 & 4.12E-06 & 1.00E-100 & 2.10E-09 & 4.50E-90 & 3.28E-40 & 6.56E-72\tabularnewline
	2.0 & 0.40 & 7.80E-06 & 1.00E-100 & 3.47E-06 & 1.52E-72 & 1.49E-09 & 5.11E-49 & 1.68E-24 & 1.24E-39\tabularnewline
	2.0 & 0.70 & 8.99E-06 & 1.49E-63 & 5.73E-06 & 6.71E-41 & 1.57E-09 & 2.07E-28 & 2.02E-17 & 3.18E-25\tabularnewline
\bottomrule
\end{tabular}}
\end{table}

\begin{table}[pt]
\caption{Same as Table~\ref{Gamma-V43-46}, but for
$^{51-54}$V isotopes.}\label{Gamma-V51-54} \centering {\tiny
\begin{tabular}{cccccccccccc}
	& & & & & & & & &\\
	\toprule \multirow{2}{*}{$\log\rho$Y$_{e}$} &
	\multirow{2}{*}{T$_{9}$} & \multicolumn{2}{c}{$^{51}$V}&
	\multicolumn{2}{c}{$^{52}$V} & \multicolumn{2}{c}{$^{53}$V}&
	\multicolumn{2}{c}{$^{54}$V}\\
	\cmidrule{3-4}  \cmidrule{5-6}  \cmidrule{7-8} \cmidrule{9-10}& &
	\multicolumn{1}{c}{$\lambda_{\gamma}^{\text{ec}+\text{pe}}$} &
	\multicolumn{1}{c}{$\lambda_{\gamma}^{\text{pc}+\text{ee}}$} &
	\multicolumn{1}{c}{$\lambda_{\gamma}^{\text{ec}+\text{pe}}$} &
	\multicolumn{1}{c}{$\lambda_{\gamma}^{\text{pc}+\text{ee}}$} &
	\multicolumn{1}{c}{$\lambda_{\gamma}^{\text{ec}+\text{pe}}$} &
	\multicolumn{1}{c}{$\lambda_{\gamma}^{\text{pc}+\text{ee}}$} &
	\multicolumn{1}{c}{$\lambda_{\gamma}^{\text{ec}+\text{pe}}$} &
	\multicolumn{1}{c}{$\lambda_{\gamma}^{\text{pc}+\text{ee}}$} \\
	\midrule
	2.0 & 0.01 & 1.00E-100 & 1.00E-100 & 1.00E-100 & 1.01E-02 & 1.00E-100 & 2.77E-03 & 1.00E-100 & 4.91E-02\tabularnewline
	2.0 & 0.10 & 1.00E-100 & 3.33E-65 & 1.00E-100 & 1.01E-02 & 1.00E-100 & 2.77E-03 & 1.00E-100 & 4.91E-02\tabularnewline
	2.0 & 0.20 & 2.17E-78 & 1.33E-37 & 1.00E-100 & 1.01E-02 & 1.00E-100 & 2.77E-03 & 1.00E-100 & 4.91E-02\tabularnewline
	2.0 & 0.40 & 2.72E-44 & 4.49E-21 & 1.11E-74 & 1.01E-02 & 8.83E-77 & 2.80E-03 & 3.87E-91 & 4.91E-02\tabularnewline
	2.0 & 0.70 & 1.79E-29 & 2.00E-13 & 9.42E-45 & 9.93E-03 & 1.12E-47 & 2.84E-03 & 1.17E-54 & 5.01E-02\tabularnewline
\bottomrule
\end{tabular}}
\end{table}

\begin{table}[pt]
\caption{ Same as Table~\ref{Gamma-V43-46}, but for
$^{55-58}$V isotopes.}\label{Gamma-V55-58} \centering {\tiny
\begin{tabular}{cccccccccccc}
	& & & & & & & & &\\
	\toprule \multirow{2}{*}{$\log\rho$Y$_{e}$} &
	\multirow{2}{*}{T$_{9}$} & \multicolumn{2}{c}{$^{55}$V}&
	\multicolumn{2}{c}{$^{56}$V} & \multicolumn{2}{c}{$^{57}$V}&
	\multicolumn{2}{c}{$^{58}$V}\\
	\cmidrule{3-4}  \cmidrule{5-6}  \cmidrule{7-8} \cmidrule{9-10}& &
	\multicolumn{1}{c}{$\lambda_{\gamma}^{\text{ec}+\text{pe}}$} &
	\multicolumn{1}{c}{$\lambda_{\gamma}^{\text{pc}+\text{ee}}$} &
	\multicolumn{1}{c}{$\lambda_{\gamma}^{\text{ec}+\text{pe}}$} &
	\multicolumn{1}{c}{$\lambda_{\gamma}^{\text{pc}+\text{ee}}$} &
	\multicolumn{1}{c}{$\lambda_{\gamma}^{\text{ec}+\text{pe}}$} &
	\multicolumn{1}{c}{$\lambda_{\gamma}^{\text{pc}+\text{ee}}$} &
	\multicolumn{1}{c}{$\lambda_{\gamma}^{\text{ec}+\text{pe}}$} &
	\multicolumn{1}{c}{$\lambda_{\gamma}^{\text{pc}+\text{ee}}$} \\
	\midrule
	2.0 & 0.01 & 1.00E-100 & 2.09E-02 & 1.00E-100 & 1.17E+01 & 1.00E-100 & 7.59E-01 & 1.00E-100 & 1.22E+01\tabularnewline
	2.0 & 0.10 & 1.00E-100 & 1.93E-02 & 1.00E-100 & 1.17E+01 & 1.00E-100 & 7.01E-01 & 1.00E-100 & 1.22E+01\tabularnewline
	2.0 & 0.20 & 1.00E-100 & 1.84E-02 & 1.00E-100 & 1.16E+01 & 1.00E-100 & 6.82E-01 & 1.00E-100 & 1.22E+01\tabularnewline
	2.0 & 0.40 & 1.00E-100 & 1.79E-02 & 1.49E-95 & 1.08E+01 & 1.00E-100 & 6.73E-01 & 1.00E-100 & 1.23E+01\tabularnewline
	2.0 & 0.70 & 1.02E-64 & 1.76E-02 & 1.16E-57 & 1.01E+01 & 1.15E-83 & 6.68E-01 & 5.78E-75 & 1.48E+01\tabularnewline
\bottomrule
\end{tabular}}
\end{table}

\begin{table}[pt]
\caption{\small Same as Table~\ref{Gamma-V43-46}, but for
$^{59-61}$V isotopes.}\label{Gamma-V59-61} \centering {\tiny
\begin{tabular}{cccccccccc}
	& & & & & & & & &\\
	\toprule \multirow{2}{*}{$\log\rho$Y$_{e}$} &
	\multirow{2}{*}{T$_{9}$} & \multicolumn{2}{c}{$^{59}$V}&
	\multicolumn{2}{c}{$^{60}$V} & \multicolumn{2}{c}{$^{61}$V}\\
	\cmidrule{3-4}  \cmidrule{5-6}  \cmidrule{7-8} & &
	\multicolumn{1}{c}{$\lambda_{\gamma}^{\text{ec}+\text{pe}}$} &
	\multicolumn{1}{c}{$\lambda_{\gamma}^{\text{pc}+\text{ee}}$} &
	\multicolumn{1}{c}{$\lambda_{\gamma}^{\text{ec}+\text{pe}}$} &
	\multicolumn{1}{c}{$\lambda_{\gamma}^{\text{pc}+\text{ee}}$} &
	\multicolumn{1}{c}{$\lambda_{\gamma}^{\text{ec}+\text{pe}}$} &
	\multicolumn{1}{c}{$\lambda_{\gamma}^{\text{pc}+\text{ee}}$} \\
	\midrule
	2.0 & 0.01 & 1.00E-100 & 5.75E+00 & 1.00E-100 & 2.04E+01 & 1.00E-100 & 3.78E+01\tabularnewline
	2.0 & 0.10 & 1.00E-100 & 5.74E+00 & 1.00E-100 & 2.04E+01 & 1.00E-100 & 3.78E+01\tabularnewline
	2.0 & 0.20 & 1.00E-100 & 5.70E+00 & 1.00E-100 & 2.04E+01 & 1.00E-100 & 3.78E+01\tabularnewline
	2.0 & 0.40 & 1.00E-100 & 5.43E+00 & 1.00E-100 & 2.04E+01 & 1.00E-100 & 3.78E+01\tabularnewline
	2.0 & 0.70 & 2.21E-96 & 5.13E+00 & 1.17E-85 & 2.06E+01 & 1.00E-100 & 3.79E+01\tabularnewline
\bottomrule
\end{tabular}}
\end{table}
\clearpage
\begin{table}[pt]
\caption{\small Same as Table~\ref{Gamma-V43-46}, but for
$^{62-64}$V isotopes.}\label{Gamma-V62-64} \centering {\tiny
\begin{tabular}{cccccccccc}
	& & & & & & & & &\\
	\toprule \multirow{2}{*}{$\log\rho$Y$_{e}$} &
	\multirow{2}{*}{T$_{9}$} & \multicolumn{2}{c}{$^{62}$V}&
	\multicolumn{2}{c}{$^{63}$V} & \multicolumn{2}{c}{$^{64}$V}\\
	\cmidrule{3-4}  \cmidrule{5-6}  \cmidrule{7-8} & &
	\multicolumn{1}{c}{$\lambda_{\gamma}^{\text{ec}+\text{pe}}$} &
	\multicolumn{1}{c}{$\lambda_{\gamma}^{\text{pc}+\text{ee}}$} &
	\multicolumn{1}{c}{$\lambda_{\gamma}^{\text{ec}+\text{pe}}$} &
	\multicolumn{1}{c}{$\lambda_{\gamma}^{\text{pc}+\text{ee}}$} &
	\multicolumn{1}{c}{$\lambda_{\gamma}^{\text{ec}+\text{pe}}$} &
	\multicolumn{1}{c}{$\lambda_{\gamma}^{\text{pc}+\text{ee}}$} \\
	\midrule
	2.0 & 0.01 & 1.00E-100 & 8.07E+01 & 1.00E-100 & 1.17E+02 & 1.00E-100 & 1.56E+01\tabularnewline
	2.0 & 0.10 & 1.00E-100 & 8.07E+01 & 1.00E-100 & 1.17E+02 & 1.00E-100 & 1.56E+01\tabularnewline
	2.0 & 0.20 & 1.00E-100 & 8.09E+01 & 1.00E-100 & 1.18E+02 & 1.00E-100 & 1.56E+01\tabularnewline
	2.0 & 0.40 & 1.00E-100 & 9.51E+01 & 1.00E-100 & 1.31E+02 & 1.00E-100 & 1.56E+01\tabularnewline
	2.0 & 0.70 & 4.12E-100 & 1.57E+02 & 1.00E-100 & 1.59E+02 & 1.00E-100 & 1.63E+01\tabularnewline
\bottomrule
\end{tabular}}
\end{table}

\end{document}